\pdfoutput=1

\documentclass[aps,prlonecolumn]{revtex4-1}
\usepackage{geometry}
\usepackage{mathptmx}
\usepackage{epsfig,amsopn}
\usepackage{graphicx}
\usepackage{graphicx}
\usepackage{amsmath,amssymb}
\usepackage{natbib}
\usepackage{xcolor}
\usepackage{braket}
\usepackage{float}
\usepackage{enumerate}
\usepackage{mathtools}
\usepackage[symbol]{footmisc}
\allowdisplaybreaks[1]

\newcommand{\eref}[1]{Eq.~(\ref{#1})}%

\def\bea{\begin{eqnarray}}
\def\eea{\end{eqnarray}}

\begin{document}

\title{{\Large{}Occupancy correlations in the asymmetric simple inclusion process}}

\author{{\normalsize{}Ofek Lauber Bonomo$^1$}{\normalsize{}}}

\author{{\normalsize{}Shlomi Reuveni$^1$}{\normalsize{}}}
\email{Author for correspondence: shlomire@tauex.tau.ac.il}

\affiliation{\noindent \textit{
$^{1}$School of Chemistry, The Center for Physics and Chemistry of Living Systems, The Raymond and Beverly Sackler Center for Computational Molecular and Materials Science,\\ \& The Mark Ratner Institute for Single Molecule Chemistry, Tel Aviv University, Tel Aviv 6997801, Israel\\}}

\date{\today}

\begin{abstract}
\noindent The asymmetric simple inclusion process (ASIP) --- a lattice-gas model for unidirectional transport with irreversible aggregation --- has been proposed as an inclusion counterpart of the asymmetric simple exclusion process and as a batch service counterpart of the tandem Jackson network. To date, analytical tractability of the model has been limited: while the average particle density in the model is easy to compute, very little is known about its joint occupancy distribution. To partially bridge this gap, we study occupancy correlations in the ASIP. We take an analytical approach to this problem and derive an exact formula for the covariance matrix of the steady-state occupancy vector. We verify the validity of this formula numerically in small ASIP systems, where Monte-Carlo simulations can provide reliable estimates for correlations in reasonable time, and further use it to draw a comprehensive picture of spatial occupancy correlations in ASIP systems of arbitrary size.
\end{abstract}
\maketitle

\section{INTRODUCTION}

\noindent Many complex and fundamental processes in nature incorporate a high level of intrinsic randomness. Stochastic events stand at the very bedrock of their micro-level description, and the cumulative effect of these events is manifested in their dynamics and functionality. Complex processes may sometimes be viewed as an interconnected network whose basic building blocks are processes of diminished complexity. Interestingly, even when the isolated behavior of each building block is understood in great detail, the behavior of the aggregate can often be extremely hard to predict. A subclass of systems in which considerable progress has, nevertheless, been made is collectively referred to as Tandem Stochastic Systems (TSS) --- linear stochastic networks formed by the sequential concatenation of relatively simple building blocks. The prevalence of TSS throughout the sciences renders this particular class of stochastic networks a special case of interest among a host of scientific communities. 

Tandem stochastic systems are systems in which a stochastic input flow (of jobs, molecules, particles, etc.) progresses through a serial array of sites. The progress from one site to the next is governed by a set of rules characterizing the system's law of motion. TSS naturally emerge in many scientific fields, including biology, chemistry, physics, and operations research, and often exhibit complex stochastic dynamics. The existing body of knowledge on TSS comprises of a small number of mathematical models that were introduced throughout the years. Each of these models was tailored to describe a different tandem stochastic system. Facing the significant complexity that is inherent to the analysis of TSS in general, this somewhat ad-hoc approach is still considered inevitable. 

This paper presents new results for  occupancy correlations in the Asymmetric Simple Inclusion Process (ASIP) --- a TSS which serves as a lattice-gas model for unidirectional transport with irreversible aggregation \cite{ASIP-1,ASIP-2}. Before coming to the ASIP, we will provide a brief overview of two other TSS: (i) the Tandem Jackson Network (TJN); and (ii) the Asymmetric Simple Exclusion Process (ASEP). These well known TSS are strongly related to the ASIP and reviewing them will help put the findings presented here in context.  

\subsection{\label{subsec:The-Tandem-Jackson}The Tandem Jackson Network}

\noindent The Tandem Jackson Network (TJN) is one of the first TSS ever to be studied. It was introduced and analyzed by R.R.P. Jackson when he was working for the operational research branch of the London airport \cite{R. R. P. Jackson 1,R. R. P. Jackson 2}. Jackson was inspired by a visit to a factory in which aircraft engines were overhauled. As he explains in the introduction to his paper:

\medskip

\hfill\begin{minipage}{\dimexpr\textwidth-0.5in}
\emph{``Work was carried out on the engines in successive stages, e.g. stripping, detailed examination, repairs, assembly and testing, and thus engines could experience inter-phase queueing. It was thought that a mathematical investigation into such a system would be helpful in planning future work and increasing the present efficiency.} [An apology was then quick to come] \emph{Unfortunately, to date, the only system of this type which appears to be mathematically tractable is one which is completely random in character, and this is now investigated."}
\end{minipage}

\medskip

\noindent In its simplest version, the TJN is a sequential array of $n$ service stations, where external jobs arrive at the leftmost station randomly in time and progress sequentially from station to station. At each station: (i) arriving jobs queue up in line and await service; (ii) only one job is served at a time; (iii) service durations are random and come from an exponential distribution; and (iv) upon completion of service: a single job moves on to the next station or, in the case of the rightmost station, out of the system. 

In the standard Queueing theory setting, the TJN can be described
as a sequential array of Markovian queues \cite{Harchol-Balter}. Jobs arrive to the first queue according to a Poisson process with rate $\lambda$ and are processed, one by one and according to order of arrival, with rate $\mu_{k}$ at station $k$. Denoting the number of jobs present in the $k^{th}$ station ($k=1,\cdots,n$) by $X_{k}$, the TJN's dynamics can be schematically summarized as follows:\\

\noindent (i) first station ($k=1$) 
\begin{equation}
X_{1},X_{2},\cdots\xrightarrow{\lambda}X_{1}+1,X_{2},\cdots\,;\label{3.1}
\end{equation}
(ii) interior stations (when $X_{k}>0$, $1<k\leq n-1$) 
\begin{equation}
\cdots,X_{k-1},X_{k},X_{k+1},\cdots\xrightarrow{\mu_{k}}\cdots,X_{k-1},X_{k}-1,X_{k+1}+1,\cdots\,;\label{3.2}
\end{equation}
(iii) last station (when $X_{n}>0$)
\begin{equation}
\cdots,X_{n-1},X_{n}\xrightarrow{\mu_{n}}\,\cdots,X_{n-1},X_{n}-1\,.\label{3.3}
\end{equation}

When $n=1$, the TJN is composed of a single service station which,
using the notation introduced by Kendall \cite{Kendall}, is equivalent to a simple $M/M/1$ queue. When $\lambda<\mu_1$, this queue is stable and its steady-state distribution is given by \cite{Harchol-Balter} 
\begin{equation}
Pr(X_{1}=x_{1})=(1-\rho_{1})\rho_{1}^{x_{1}}\,,\label{3.4}
\end{equation}
where $\rho_{1}=\lambda/\mu_{1}$ $(x_{1}=0,1,2,\cdots$). While this result was already known to Jackson and his contemporaries, its extension to $n>1$ stations was considered nontrivial. Indeed, when the output from one station is the input of the next, intricate correlations may (in principle) arise and render the joint probability distribution of the system extremely complex or even completely intractable.  

In light of the above, it is quite remarkable that the steady-state distribution of a TJN with $n$ service stations has the following product form  
\begin{equation}
Pr(X_{1}=x_{1},\cdots,X_{n}=x_{n})=\overset{n}{\underset{k=1}{\prod}}(1-\rho_{k})\rho_{k}^{x_{k}}\,,\label{3.5}
\end{equation}
provided that $\rho_{k}=\lambda/\mu_{k}<1$ $(x_{k}=0,1,2,\cdots)$. Equation
(\ref{3.5}) is known as Jackson's theorem, it asserts that stations in the TJN behave as if they were a collection of $n$ separate $M/M/1$ queues that are statistically independent of each other. The great simplicity and elegance of this result is best appreciated when compared to the two other TSS to be described hereinafter.

\subsection{The Asymmetric Simple Exclusion Process}

\noindent The Asymmetric Simple Exclusion Process (ASEP) --- a stochastic process taking place on a discrete one-dimensional lattice of $n$ sites --- is a paradigmatic model in non-equilibrium statistical physics \cite{Derrida1,Golinelli,Derrida2,Blythe}. Having first appeared in the literature as a model of bio-polymerization \cite{MacDonald}, it was later introduced to the probability theory and statistical-physics communities by Frank Spitzer \cite{Spitzer}, and has now become a default model for stochastic transport with excluded volume interactions. Over the years, the ASEP and its variants were used to study a wide range of physical phenomena: transport across membranes \cite{Heckmann}, transport of macromolecules through thin vessels \cite{Levitt}, hopping conductivity in solid electrolytes \cite{Richards}, reptation of a polymer in a gel \cite{Widom}, traffic flow \cite{Schreckenberg}, protein synthesis by ribosomes \cite{Shaw,Reuveni}, surface growth \cite{Halpin,Krug}, sequence alignment \cite{Bundschuh}, molecular motors \cite{Klumpp} and the directed motion of tracer particles in the presence of dynamical backgrounds \cite{Oshanin1,Oshanin2,Oshanin3,Monasterio}.

In the ASEP, particles flow randomly in time, into the leftmost site of a one-dimensional lattice and propagate unidirectionally (to the right) along the lattice. Particles move from a site to its right-neighboring site randomly in time --- the hopping restricted by the exclusion principle which allows sites to be occupied by no more than one particle at a time. At the rightmost site, particles exit the system randomly in time. The exclusion principle causes jamming throughout the lattice, and renders the ASEP's dynamics highly non-trivial.

The translation between the queueing-theory setting of the TJN and the statistical-physics setting of the ASEP is straightforward: ‘jobs’ are ‘particles’ and ‘service stations’ are ‘sites’. The random inflow into the leftmost site in the ASEP, the random instants of hopping from site to site, and the random outflow from the rightmost site are all governed by independent Poisson processes. Denoting an occupied site by $\bullet$ and an empty site by $\circ$, the ASEP's dynamics can be schematically summarized as follows:
\\

\noindent (i) first site ($k=1$)
\begin{equation}
\circ,\cdots\,\xrightarrow{\lambda}\,\bullet,\cdots\,;\label{3.6}
\end{equation}
(ii) interior sites $(1<k\leq n-1)$
\begin{equation}
\cdots,\bullet,\circ,\cdots\xrightarrow{\mu_{k}}\cdots,\circ,\bullet,\cdots\,;\label{3.7}
\end{equation}
(iii) last site ($k=n$)
\begin{equation}
\cdots,\bullet\,\xrightarrow{\mu_{n}}\,\cdots,\circ\,.\label{3.8}
\end{equation}
Recalling that the capacity of a queue is the maximum number of jobs allowed in it (including those in service); One can think of the ASEP as a TJN of Markovian queues with single job capacity. When this number is reached, further arrivals to the first site are turned away and blocking occurs in interior sites. 

In contrast to the steady-state distribution of the TJN, that of the ASEP does not obey a product form, i.e., particle occupancies in distinct sites are statistically dependent. It is nevertheless interesting to note that exact expressions for the steady-state distribution can sometimes be put in the form of a matrix-product. To this end, the statistical weight of each of the $2^{n}$ possible configurations of an ASEP lattice is constructed as a product of matrices, one for each site, chosen according to the state of the site (occupied or empty). The probability to observe the lattice in a particular configuration can then be obtained following proper normalization. A matrix-product solution for the steady-state distribution of the ASEP was first derived in \cite{Derrida3}. Matrix-product forms are reviewed in \cite{Blythe}.

\subsection{The Asymmetric Simple Inclusion Process}

\noindent Exclusion is central to the ASEP, but while this principle is suitable for the description of some physical systems it is not suitable for others. Altering the ASEP such that arbitrarily many particles are allowed to simultaneously occupy any given site, one can end up with two different models: the tandem Jackson network that was discussed above, and the Asymmetric Simple Inclusion Process (ASIP) --- a model which was introduced in \cite{ASIP-1,ASIP-2} and further analysed in \cite{ASIP-3,ASIP-4,ASIP-5}. The ASIP is similar to the ASEP --- albeit replacing the exclusion principle by an inclusion principle. In both processes, random events cause particles to propagate unidirectionally along a one-dimensional lattice. In the ASEP particles are subject to \emph{exclusion} interactions that keep them singled apart, whereas in the ASIP particles are subject to \emph{inclusion} interactions that coalesce them into inseparable clusters. To avoid confusion we note that there is also a different model with the name ASIP \cite{Not-ASIP1,Not-ASIP2} which will not be discussed here. 

The formulation of the ASIP is as follows. Consider a one-dimensional lattice of $n$ sites indexed $k=1,\cdots,n$. Each site is followed by a gate --- labeled by the site's index --- which controls the site's outflow. Particles arrive at the first site ($k=1)$ following a Poisson process with rate $\lambda$, the openings of gate $k$ are timed according to a Poisson process with rate $\mu_{k}$ ($k=1,\cdots,n$), and the $n+1$ Poisson processes are mutually independent. A key feature of the ASIP is its `batch service' property: at an opening of gate $k$ all particles present at site $k$ transit simultaneously, and in one batch (one cluster), to site $k+1$ --- thus joining particles that may already be present at site $k+1$ ($k=1,\cdots,n-1$). At an opening of the last gate ($k=n$) all particles present at site $n$ exit the lattice simultaneously. Denoting the number of particles present in site $k$ ($k=1,\cdots,n$) by $X_{k}$, the ASIP's dynamics can be schematically summarized as follows:\\

\noindent (i) first site ($k=1$)
\begin{equation}
X_{1},X_{2},\cdots\xrightarrow{\lambda}X_{1}+1,X_{2},\cdots\,;\label{3.9}
\end{equation}
(ii) interior sites $(1<k\leq n-1)$
\begin{equation}
\cdots,X_{k-1},X_{k},X_{k+1},\cdots\xrightarrow{\mu_{k}}\cdots,X_{k-1},0,X_{k+1}+X_{k},\cdots\,;\label{3.10}
\end{equation}
(iii) last site ($k=n$)
\begin{equation}
\cdots,X_{n-1},X_{n}\xrightarrow{\mu_{n}}\,\cdots,X_{n-1},0\,.\label{3.11}
\end{equation}

From a queueing theory perspective, the ASIP is a sequential array of Markovian queues with unbounded capacity and unlimited batch service \cite{Neuts,Kaspi}: all particles present at a given service station are served collectively (and thus move together to the next service station or out of the system). The notion of ‘batch service’ is strongly related to growth-collapse processes. Consider a single service station with batch service. Jobs arrive to the station randomly in time --- causing the queue to grow steadily; when service is rendered all jobs are served simultaneously --- causing the queue to collapse to zero. Thus, stochastic growth-collapse temporal patterns emerge from the application of batch-service policies \cite{Neuts,Kaspi,Boxma,Kella,Martin}. Interestingly, these patterns also appear in a variety of complex systems, including: sand-pile models and systems in self-organized criticality \cite{sandpile}, stick-slip models of interfacial friction \cite{Rozman}, Burridge-Knopoff type models of earthquakes and continental drift \cite{Carlson}, stochastic avalanche models \cite{Eliazar1}, stochastic Ornstein-Uhlenbeck capacitors \cite{Eliazar2} and geometric Langevin equations \cite{Eliazar3}. The ASIP model is, in effect, a tandem array of growth-collapse processes. 

From a statistical physics perspective, the ASIP is a model for unidirectional transport with irreversible aggregation/coagulation. Such reaction-diffusion models have been extensively studied since the pioneering work of Smoluchowski \cite{Smoluchowski}. Yet still, unresolved issues and intriguing new facets cause them to raise interest even today \cite{Sokolov,Lindenberg}. Two of the simplest models of this kind are the coalescence-diffusion model 
\begin{align}
\cdots\bullet\bullet\cdots & \xrightarrow{1}\,\cdots\circ\bullet\cdots\;,\nonumber \\
\cdots\bullet\circ\cdots & \xrightarrow{1}\,\cdots\circ\bullet\cdots\;,\label{3.12}
\end{align}
where $\bullet$ represents an occupied site and $\circ$ represents
an empty site, and the aggregation-diffusion model
\begin{align}
\cdots A_{l}A_{l'}\cdots & \xrightarrow{1}\,\cdots0A_{l+l'}\cdots\;,\nonumber \\
\cdots A_{l}0\cdots & \xrightarrow{1}\,\cdots0A_{l}\cdots\;,\label{3.12-1}
\end{align}
where $A_{l}$ represents a site occupied by $l>0$ particles, and
$0$ represents an empty site \cite{Kinetic View}. 

The studies dedicated to the models described in Eqs. (\ref{3.12}) and (\ref{3.12-1}) were, by and large, carried out in a one-dimensional ring topology. Under these conditions, many statistical properties can be calculated exactly using the empty-interval method \cite{Kinetic View,Coalescence Process}. The ASIP, with homogeneous unit rates $\{\mu_{1}=\cdots=\mu_{n}=1\}$, can be viewed as a generalization of aggregation-diffusion models to an open system. Indeed, the bulk ASIP dynamics of Eq. (\ref{3.10}) is identical to the dynamics of Eq. (\ref{3.12-1}). Similarly, when one disregards the number of particles occupying each site ($X_{k}$) and focuses only on whether sites are occupied or not ($X_{k}>0$ or $X_{k}=0$), the ASIP dynamics turns into an open-boundary version of Eq. (\ref{3.12}).

\subsection{Connection between the TJN, ASEP and ASEP} 

\begin{table}
\begin{centering}
{\scriptsize{}}%
\begin{tabular}{|c|c|c|}
\hline 
 & $c_{site}=1$ & $c_{site}=\infty$\tabularnewline
\hline 
\hline 
$c_{gate}=1$ & {\scriptsize{}ASEP} & {\scriptsize{}TJN}\tabularnewline
\hline 
$c_{gate}=\infty$ & {\scriptsize{}ASEP} & {\scriptsize{}ASIP}\tabularnewline
\hline 
\end{tabular}{\scriptsize\par}
\par\end{centering}
\caption{\label{Table3.1}Capacity classification of the TJN, ASEP, and ASIP
models.}
\end{table}

\noindent Interestingly, all three models --- TJN, ASEP, and ASIP --- can be described by the sites-gates language that was used to describe the ASIP. To pinpoint the difference between the models consider the two following characteristic capacities: (i) \emph{site capacity} $c_{site}$ --- the maximal number of particles that can simultaneously occupy a given site, and (ii) \emph{gate capacity} $c_{gate}$ --- the maximal number of particles that are simultaneously transferred through a given gate when it opens. In each of the above-mentioned models particles propagate according to the following rule: at an opening of gate $k$, $min(X_{k},c_{site}-X_{k+1},c_{gate})$ particles transit simultaneously from site $k$ to site $k+1$ --- thus joining particles that may already be present at site $k$ ($k=1,2,\cdots,n-1$). At an opening of the last gate $(k=n)$, $min(X_{k},c_{gate})$ particles exit the lattice simultaneously. In the ASEP the site capacity is $c_{site}=1$ and the gate capacity can be any positive integer $1\leq c_{gate}\leq\infty$. In the TJN the site capacity is $c_{site}=\infty$ and the gate capacity is $c_{gate}=1$. In the ASIP the site capacity is $c_{site}=\infty$ and the gate capacity is $c_{gate}=\infty$. The capacity classification is summarized in Table \ref{Table3.1} --- from which it is evident that the ASIP is, in effect, a ‘missing puzzle piece’ connecting together the well established and the well studied ASEP and TJN models.

\subsection{Asymmetric Simple Inclusion Process: Fundamental Results} 

\noindent Many fundamental results that were obtained for the ASIP made use of its Markovian dynamics which we now describe. Set $\mu_{c}=\mu_{1}+\cdots+\mu_{n}$ to be the ASIP's cumulative hopping rate, let $\mathrm{X}_{\mathrm{k}}\left(t\right)$ denote the number of particles present in the $k^{\text{th}}$ site ($k=1,\ldots,n$) at time $t$ ($t\geq0$), and set $\mathbf{X}\left(n,t\right)=\left(\mathrm{X_{1}}\left(t\right),\cdots,\mathrm{X_{n}}\left(t\right)\right)^{{\color{brown}\mathrm{{\color{blue}{\color{black}\top}}}}}.$ The vector $\mathbf{X}\left(n,t\right)$ represents the occupancy of an ASIP system with $n$ sites at time $t$. Observe the system at times $t$ and $t^{\prime}=t+\Delta$ (for small $\Delta$) and use the shorthand notation $\mathbf{X}=\mathbf{X}\left(n,t\right)$ and $\mathbf{X}^{\prime}=\mathbf{X}\left(n,t^{\prime}\right)$. The stochastic connection between the random vectors $\mathbf{X}$ and $\mathbf{X}^{\prime}$ characterizing the Markovian `law of motion' of the stochastic process $[\mathbf{X}\left(t\right)]_{t\geq0}$ is given by:
\textcolor{black}{
\begin{equation}
\left(\mathrm{X_{1}^{\prime},\cdots,X_{n}^{\prime}}\right)=\left\{ \begin{array}{lll}
\left(\mathrm{X_{1},X_{2},X_{3},\cdots,X_{n-1},X_{n}}\right) & \text{ \ } & \text{w.p. }1-\left(\lambda+\mu_{c}\right)\Delta+o(\Delta)\text{ ,}\\
\text{ \ }\\
\left(\mathrm{X_{1}+1,X_{2},X_{3},\cdots,X_{n-1},X_{n}}\right) &  & \text{w.p. }\lambda\Delta+o(\Delta)\text{ ,}\\
\text{ \ }\\
\left(\mathrm{0,X_{1}+X_{2},X_{3},\cdots,X_{n-1},X_{n}}\right) &  & \text{w.p. }\mu_{1}\Delta+o(\Delta)\text{ ,}\\
\text{ \ \ }\\
\left(\mathrm{X_{1},0,X_{2}+X_{3},\cdots,X_{n-1},X_{n}}\right) &  & \text{w.p. }\mu_{2}\Delta+o(\Delta)\text{ ,}\\
\vdots &  & \vdots\\
\mathrm{\left(X_{1},X_{2},X_{3},\cdots,0,X_{n-1}+X_{n}\right)} &  & \mathrm{\text{w.p. }\mu_{n-1}\Delta+o}(\Delta)\text{ ,}\\
\text{ \ \ }\\
\left(\mathrm{X_{1},X_{2},X_{3},\cdots,X_{n-1},0}\right) & \text{ \ \ } & \text{w.p. }\mathrm{\mu_{n}}\Delta+o(\Delta)\text{ ,}
\end{array}\right.\label{markov dynamics}
\end{equation}} where w.p. is short for "with probability".  

Equation \ref{markov dynamics} follows
from considering the totality of events that may take place within
the time interval $(t,t^{\prime}]$: ($0$) arrival of a particle
to the first site occurring with probability $\lambda\Delta+o(\Delta)$ in which case $X_{1}\mapsto X_{1}^{\prime}=X_{1}+1$; ($1$) transition of all particles from the first site to the second site occurring with probability $\mu_{1}\Delta+o(\Delta)$ in which case $X_{1}\mapsto X_{1}^{\prime}=0$ and $X_{2}\mapsto X_{2}^{\prime}=X_{1}+X_{2}$; ($2$) transition of all particles from the second site to the third site occurring with probability $\mu_{2}\Delta+o(\Delta)$ in which case $X_{2}\mapsto X_{2}^{\prime}=0$ and $X_{3}\mapsto X_{3}^{\prime}=X_{2}+X_{3}$; $\cdots$; ($n-1$) transition of all particles from the site before last to the last site occurring with probability $\mu_{n-1}\Delta+o(\Delta)$ in which case $X_{n-1}\mapsto X_{n-1}^{\prime}=0$ and $X_{n}\mapsto X_{n}^{\prime}=X_{n-1}+X_{n}$; ($n$) transition of all particles from the last site to outside of the lattice occurring
with probability $\mu_{n}\Delta+o(\Delta)$ in which case $X_{n}\mapsto X_{n}^{\prime}=0$. The first line on the right-hand-side of Eq. (\ref{markov dynamics}) represents the scenario in which no
event takes place which occurs with the complementary probability $1-\left(\lambda+\mu_{c}\right)\Delta+o(\Delta)$.

In steady-state, the stochastic process $[\mathbf{X}\left(t\right)]_{t\geq0}$ is stationary, hence the means are time homogeneous: $\left\langle \mathbf{X}\left(n,t\right)\right\rangle \equiv\left\langle \mathbf{X}\left(n\right)\right\rangle \;(t\geq0).$ In \cite{ASIP-1} it was shown that the mean occupancy of the $k^{th}$ site at steady-state is given by 
\begin{equation}
{\color{black}\left\langle \mathbf{\mathrm{X}_{\mathrm{k}}}\right\rangle =\lambda/\mu_{k}}.\label{mean occupation steady state}
\end{equation}
The multidimensional probability generating function (PGF) of the occupancy vector was also studied in \cite{ASIP-1}, and explicit steady-state PGFs were obtained for small ASIP systems $(n=1,2,3)$. This made clear that similar to the ASEP \cite{Derrida1,Golinelli,Derrida2,Blythe}, and in contrast with other lattice gas models like the TJN and the related zero range process \cite{zero-range}, PGFs in the ASIP do not admit a product form. In \cite{ASIP-1}, an iterative algorithm that allows computation of steady-state PGFs for ASIP systems of arbitrary size was presented. However, explicit formulas for the PGFs of small ASIP systems are already very cumbersome and formulas for the PGFs of large ASIP systems are hard to obtain and virtually impossible to work with. 

A little more is known about homogeneous ASIP systems \cite{ASIP-1,ASIP-2}. In this subclass of ASIPs, all hopping rates are identical: ${\color{black}\mu=\mu_{\mathbf{\mathrm{1}}}=\ldots=\mu_{\mathrm{n}}}$; and the marginals of the joint occupancy distribution at steady-state can be computed exactly \cite{ASIP-4} and written in terms of Catalan numbers and their generalizations \cite{Catalan's Trapezoids}. A formula for the factorial moments of the  steady-state occupancies follows. Using this formula, it can be shown that the variance of in the occupancy of the $k^{th}$ site is given by (Appendix A)
\begin{equation}
{\color{black}\mathbf{Var[\mathbf{\mathrm{\mathrm{X}}_{\mathrm{k}}}]=}}{\color{black}\rho+\left(\frac{4\Gamma(k+1/2)}{\sqrt{\pi}\Gamma(k)}-1\right)\rho^{2}}.\label{variance}
\end{equation}
where $\rho=\lambda / \mu$ is the mean site occupancy (also average particle density) in the ASIP. In this paper, we will generalize this formula to capture the full variance-covariance matrix of homogeneous and non-homogeneous ASIPs at steady-state.  
\subsection{Focus and structure of this paper}
\noindent Occupancy correlations in the ASIP are captured by an n-dimensional variance-covariance matrix
\begin{equation}
{\color{black}\mathbf{Cov}\left[\mathbf{X}(n,t)\right]\equiv\left\langle \mathbf{X}(n,t)\mathbf{X}(n,t)^{\top}\right\rangle -\Bigl\langle\mathbf{X}(n,t)\Bigr\rangle\left\langle \mathbf{X}(n,t)^{\top}\right\rangle },\label{covariance}
\end{equation}
\noindent in which the non-diagonal elements give the covariance between different sites, and the diagonal elements give the variances in the occupancy of sites. The remainder of the paper is dedicated to this variance-covariance matrix and is organized as follows. In Sec. \textcolor{blue}{{}II}, we present a formula for the covariance matrix of a general ASIP in steady-state. In addition, we give a closed-form expression  for the covariance matrix of homogeneous ASIP systems of arbitrary size. The derivation of these results is given in Sec. \textcolor{blue}{III}\textcolor{black}{{} and}\textcolor{blue}{IV} respectively. We conclude in Sec. \textcolor{blue}{V}.

\section{A SUMMARY OF KEY RESULTS ESTABLISHED IN THIS PAPER}
\noindent In this section we present a short summary of the key results established in this paper. We give an exact formula for the occupancy correlation matrix of a general ASIP system in steady-state and a closed form expression for the correlation matrix of a homogeneous ASIP system in steady-state. We start with some basic definitions and notation. 

Throughout this paper we will use the shorthand vector notation
\begin{equation}
{\color{black}\mathbf{e_{1,n}}=(1,0,\cdot\cdot\cdot,0)^{\top}} \in \mathbb{R}^{\textit{1xn}},\label{1 vector}
\end{equation}
and the following set of linear operators
\begin{equation}
{\color{black}\mathbf{\left\{ \begin{array}{l}
\mathbf{M_{1,n}X\mathrm{(\mathit{n,t})}}=\mathrm{\left(-X_{1}(\mathit{t}),X_{1}\left(\mathit{t}\right),0,\cdots,0\right)^{\top}}\mathrm{,}\\
\\
\mathbf{M_{2,n}X\mathrm{(\mathit{n,t})}}=\mathrm{\left(0,-X_{2}\left(\mathit{t}\right),X_{2}\left(\mathit{t}\right),0,\cdots,0\right)^{\top}}\mathrm{,}\\
.\\
.\\
.\\
\mathbf{M_{n-1,n}X\mathrm{(\mathit{n,t})}}=\mathrm{\left(0,\cdots,0,-X_{n-1}\left(\mathit{t}\right),X_{n-1}\left(\mathit{t}\right)\right)^{\top},}\\
\\
\mathbf{M_{n,n}X\mathrm{(\mathit{n,t})}}=\mathrm{\left(0,\cdots,0,-X_{n}\left(\mathit{t}\right)\right)^{\top}}.
\\
\end{array}\right.}}\label{M operators}
\end{equation}
The $\mathbf{\mathbf{M_{k,n}}} \in \mathbb{R}^{\textit{nxn}}$ operators in Eq. (\ref{M operators}) represent the \textit{change} in the occupancy vector that is caused due to the opening of the $k$\textsuperscript{th} gate ($k=1,\ldots,n$) in an ASIP system of $n$ sites. The $\mathbf{\mathbf{M_{k,n}}}$ operators have the following matrix form 
\begin{equation}
{\color{black}\mathbf{\mathbf{M_{k,n}\mathrm{\boldsymbol{\mathit{(i,j)}}}}}=\begin{cases}
-1 & \mathit{k=i=j}\\
 1 & \mathit{k=i-1=j}\\
 0 & \mathit{otherwise,}
\end{cases}}\label{Mk operator}
\end{equation}
for $k<n$, and 
\begin{equation}
{\color{black}\mathbf{\mathbf{M_{n,n}\mathrm{\boldsymbol{\mathit{(i,j)}}}}}=\begin{cases}
-1 & \mathit{n=i=j}\\
0 & \mathit{otherwise,}
\end{cases}}\label{Mn operator}
\end{equation}
for $k=n$. In addition, we define the linear operator 
\begin{equation}
\mathbf{P_{n}X\mathrm{(\mathit{n,t})}}=\mathrm{\left(0,\cdots,0,X_{n-1}\left(\mathit{t}\right)\right)^{\top}} \in \mathbb{R}^{\textit{nxn}},
\label{P operator}
\end{equation}
that will prove to be useful later. 

With the aid of the above, we  define an additional set of vectors and matrices:  $\mathrm{\mathbf{V}}_{\mathrm{\mathbf{k}}} \in \mathbb{R}^{\textit{1xk}}$, $\mathbf{\mathbf{R}_{\mathrm{\mathrm{\mathbf{k}}}}} \in \mathbb{R}^{\textit{kxk}}$
and $\mathbf{Q}_{\mathrm{\mathbf{k}}}\in \mathbb{R}^{\textit{(k-1)xk}}$. We start with \textcolor{black}{$\mathrm{\mathbf{V}}_{\mathrm{\mathbf{k}}}$
} that is defined as follows
\begin{equation}
{\color{black}\mathrm{\mathbf{V}}_{\mathbf{\mathbf{\mathrm{\mathbf{k}}}}}=\frac{\lambda^{2}}{\mu_{\mathrm{\textit{k}}}}\mathbf{e^{\top}_{1,k}}}. \label{V vector}
\end{equation} 
We continue with \textcolor{black}{$\mathbf{\mathbf{R}_{\mathrm{\mathrm{\mathbf{k}}}}}$} that is defined to be  
\textcolor{black}{
\begin{equation}
\mathbf{\mathbf{R}_{\mathbf{\mathbf{\mathrm{\mathbf{k}}}}}}=\overset{\mathrm{k}-1}{\underset{i=1}{\sum}}\mathrm{\mu_{\textit{i}}}\mathbf{\mathbf{M_{i,k}^{\top}}}+\mathrm{\mu_{\textit{k}-1}}\mathbf{P_{k}^{\top}}-\mathrm{\mu_{\textit{k}}}\text{\textbf{I}},\label{R matrix}
\end{equation}}
\noindent where $\mu_i$ is the gate opening rate of the i-th gate, the matrices $\mathbf{\mathbf{M_{i,k}}}$ were defined in Eq. (\ref{Mk operator}), the matrix $\mathbf{\mathbf{P_{k}}}$ was defined in Eq. (\ref{P operator}), and ${\color{black}\mathbf{I}}$ stands for the identity matrix. We observe that $\mathbf{R}_{\mathrm{\mathbf{k}}}$ is
an upper triangular matrix of the form

\begin{equation}
{\color{black}\mathbf{\mathbf{R}_{\mathrm{\mathbf{k}}}}=\left(\begin{array}{ccccc}
-\mu_{1}-\mu_{k} & \mu_{1}\\
 & -\mu_{2}-\mu_{k} & \ddots\\
 &  & \ddots & \mu_{k-2}\\
 &  &  & -\mu_{k-1}-\mu_{k} & 2\mu_{k-1}\\
 &  &  &  & -{\color{black}\mu}_{k}
\end{array}\right)}\label{R matrix explicit form}
\end{equation}
\begin{figure}[t]
\noindent \begin{centering}
\includegraphics[scale=1]{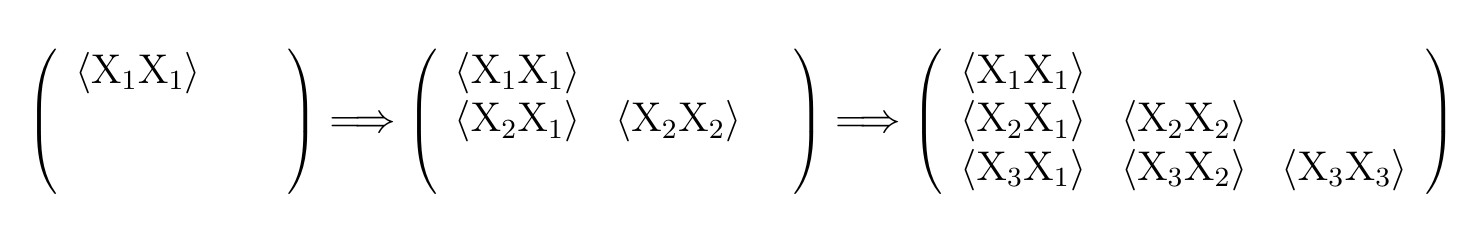}
\par \end{centering}
\caption{In section \textcolor{blue}{III}, we show that the rows of the cross-correlations matrix obey a recursive relation. Given $\left\langle \mathrm{X_{1}X_{1}}\right\rangle$ one can compute $\left\langle \mathrm{X_{2}X_{1}}\right\rangle$ and $\left\langle \mathrm{X_{2}X_{2}}\right\rangle$. Then, given $\left\langle \mathrm{X_{2}X_{1}}\right\rangle$ and $\left\langle \mathrm{X_{2}X_{2}}\right\rangle$ one can compute $\left\langle \mathrm{X_{3}X_{1}}\right\rangle$, $\left\langle \mathrm{X_{3}X_{2}}\right\rangle$ and $\left\langle \mathrm{X_{3}X_{3}}\right\rangle$. Proceeding in this way one can obtain the lower triangular half of  $\left\langle\mathbf{X}(n)\mathbf{X}(n)^{\top}\right \rangle$ and by symmetry the entire matrix.} 
\label{Fig1} 
\end{figure}
which in particular means that the inverse $\mathrm{\mathbf{\mathbf{R}_{k}^{-1}}}$ exists and is also an upper triangular matrix. Finally, we define $\mathrm{\mathbf{Q}_{\mathbf{2}}}=\begin{pmatrix}\mathrm{0},\: & \mu_{1}\end{pmatrix}$ and 
\begin{equation}
{\mathrm{{\color{black}\mathbf{Q}_{\mathbf{\mathbf{\mathrm{\mathbf{k}}}}}=\left(\begin{array}{cccccc}
{\color{black}\mu_{k-1}} & {\color{black}} & {\color{black}\cdots} & {\color{black}} &  & {\color{black}}\\
{\color{black}} & {\color{black}\mu_{k-1}} & {\color{black}} & {\color{black}} &  & {\color{black}}\\
{\color{black}\vdots} & {\color{black}} & {\color{black}\ddots} & {\color{black}} &  & {\color{black}}\\
{\color{black}} & {\color{black}} & {\color{black}} & {\color{black}\mu_{k-1}} & \vdots & {\color{black}}\\
 &  &  & \cdots & 0 & {\color{black}\mu_{k-1}}
\end{array}\right)}}}\label{Q matrices}
\end{equation}

\noindent for $k>2$. Here, and in all of the above matrices, blank spaces represent zero
entries.

We are now ready to state the main results of this paper. The covariance matrix of the occupancy vector  $\mathbf{X}\left(n\right)$ is an $nxn$ symmetric matrix that is given by
\begin{equation}
{\color{black}\mathbf{Cov}\left[\mathbf{X}(n)\right]\equiv\left\langle \mathbf{X}(n)\mathbf{X}(n)^{\top}\right\rangle -\mathbf{\Bigl\langle\mathbf{X}\mathrm{(\mathit{n})}\Bigr\rangle}\left\langle \mathbf{X}(n)^{\top}\right\rangle}. \label{covariance steady state}
\end{equation}
\noindent The second term in Eq. (\ref{covariance steady state} can be easily obtained from Eq. (\ref{mean occupation steady state}) and is given by
\begin{equation}
{\color{black}\left(\Bigl\langle\mathbf{X}(n)\Bigr\rangle\left\langle \mathbf{X}(n)^{\top}\right\rangle \right)_{\mathit{i,j}}{\color{black}{\color{black}\mathrm{=}}}\mathrm{\frac{\lambda^{2}}{\mu_{\mathit{i}}\mu_{\mathit{j}}}}}.\label{means matrix}
\end{equation}
\noindent The first term in Eq. (\ref{covariance steady state}) 
is the cross-correlations matrix $\left\langle\mathbf{X}(n)\mathbf{X}(n)^{\top}\right \rangle$. In Sec. \textcolor{blue}{III}, we show that the rows of this matrix admit a recursion relation which allows their computation (Fig. \ref{Fig1}). We start by showing that 
\begin{equation}
    \left\langle\mathrm{X_{1}X_{1}}\right\rangle=\left\langle \mathrm{X_{1}^{2}}\right\rangle =\frac{\lambda(2\lambda+\mu_{1})}{\mu_{1}^{2}}\:,
\end{equation}
and that
\begin{equation}
    \left\langle \mathrm{X}_{\mathrm{2}}\mathbf{\mathbf{X}^{\top}}(2)\right\rangle =-\mathbf{V}_{\mathbf{2}}\mathbf{\mathbf{R}_{2}^{-1}}-\left\langle\mathrm{X_{1}X_{1}}\right\rangle\mathbf{Q}_{\mathbf{2}}\mathbf{\mathbf{R}_{2}^{-1}}, 
\end{equation}
and from there continue to show that for $k \geq 3$, the $k$\textsuperscript{th} row of the lower triangular half of the cross-correlations matrix is given by 
\begin{equation}
{\color{black}\left\langle \mathrm{X}_{\mathrm{k}}\mathbf{\mathbf{X}^{\top}}(k)\right\rangle =-\mathbf{\mathbf{V}_{k}}\mathbf{\mathbf{R}_{k}^{-1}}+\sum_{\mathrm{i=2}}^{\mathrm{k-1}}(-1)^{\mathrm{k-i+1}}\mathrm{\mathbf{V}}_{\mathrm{\mathbf{i}}}\mathbf{\mathbf{R}_{\mathbf{\mathbf{\mathrm{\mathbf{i}}}}}^{-1}}\mathrm{\prod_{j=i+1}^{k}}\mathrm{\mathbf{Q}_{\mathbf{j}}}\mathbf{\mathbf{R}_{\mathrm{\mathrm{\mathbf{j}}}}^{-1}}+{\color{blue}{\color{black}(-1)^{\mathrm{k-1}}\frac{{\color{black}\lambda(2\lambda+\mu_{1})}}{{\color{black}\mu_{1}^{2}}}}}\prod_{\mathrm{i=2}}^{\mathrm{k}}\mathrm{\mathbf{Q}_{\mathrm{\mathbf{i}}}}\mathbf{\mathbf{R}_{\mathrm{\mathrm{\mathbf{i}}}}^{-1}}} .\label{general correlations}
\end{equation}
The full matrix can then be obtained by reflecting the entries of the lower triangular matrix about its main diagonal.

\begin{figure}[t]
\includegraphics[width=16cm]{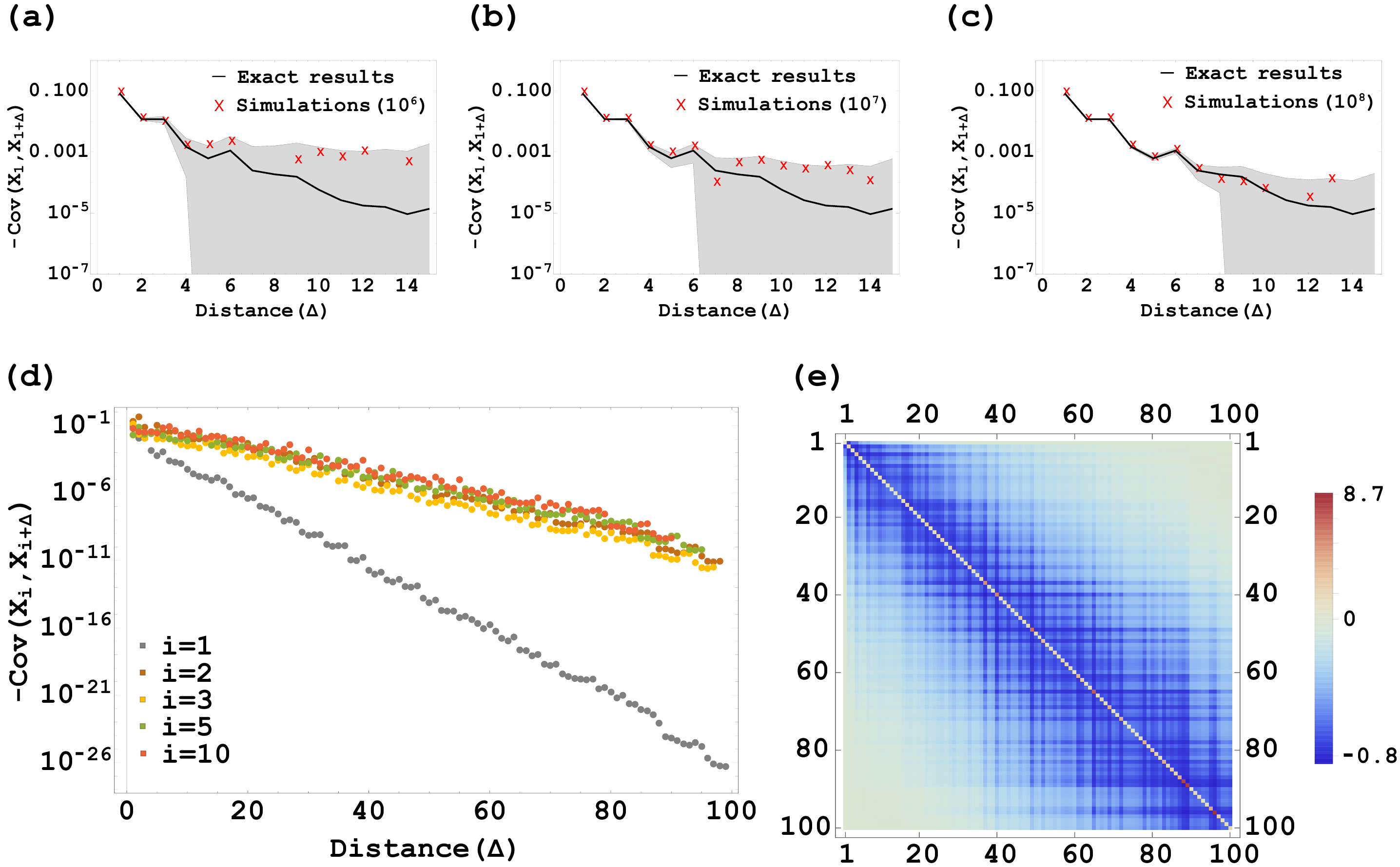}
\caption{Numerical validation and graphical illustration of the result in Eq. (\ref{general correlations}). Panels (a)-(c): comparison with numerical simulations. The exact result for the covariance between $X_1$ and $X_{1+\Delta}$ is plotted vs. the distance $\Delta$ between the two sites (solid black line). Note that the covariance is always negative and that it quickly drops with the distance. Results from simulations are presented as red crosses. These were obtained by sampling $10^6$, $10^7$ and $10^8$ occupancy vectors (panels (a), (b) and (c) respectively). One can observe that the uncertainty around the numerical estimates (shaded in gray) slowly diminishes as the number of samples increases, thus converging to the black theory line. However, it is also clear that obtaining numerical estimates for the correlation between sites that are far apart is virtually impossible in the ASIP as this would typically require a huge number of samples which cannot be obtained in reasonable time. This emphasizes the importance of exact solutions. Panel (d): The exact result for the covariance between $X_i$ and $X_{i+\Delta}$ is plotted as a function of the distance $\Delta$ for $i=1,2,3,5,10$. As expected, correlations generally become weaker with distance, but now one can also observe that the decay is exponential. Panel (e): Heat map representation of the covariance matrix $\left\langle\mathbf{X}(n)\mathbf{X}(n)^{\top}\right \rangle$. Correlations are color coded: deep red/blue represents strong positive/negative correlation and white represents weak correlation. The interesting pattern seen is due to the digits structure of the number $\pi$.}
\label{Fig2}
\end{figure} 

In Fig. \ref{Fig2}, we graphically demonstrate various aspects of the result in Eq. (\ref{general correlations}). We do this with an ASIP of $n=100$ sites in which $\lambda=1$ and the $\mu_k$'s were taken to be the first  100 non-zero digits of the number $\pi$. First, we compare our analytical result with numerical simulations (Fig. \ref{Fig2}, panels a,b,c). We observe that correlations in the ASIP model drop fast with the distance between sites which in turn requires high-precision simulations. Indeed, even when sampling $10^6$ steady-state occupancy vectors, reliable estimates could only be obtained for the covariance of the first site with its four consecutive sites (Fig. 2a), and increasing the number of samples by a factor of 10 and 100 was only able to increase the number of reliable estimates to six, and eight, correspondingly. In their range of validity, numerical simulations agree well with Eq. (\ref{general correlations}); but the hardship involved in extending this range also make it clear that long-distance correlations in the ASIP can only be obtained analytically. Panels (d) and (e) of Fig. \ref{Fig2} show that long-distance correlations can be easily obtained using Eq. (\ref{general correlations}). We observe that the negative correlation between site occupancies decays exponentially with distance (Fig. 2d). The interesting pattern on top of the general trend is due to the digits structure of the number $\pi$ (Fig. 2e). 

\begin{figure}[t]
\includegraphics[width=16cm]{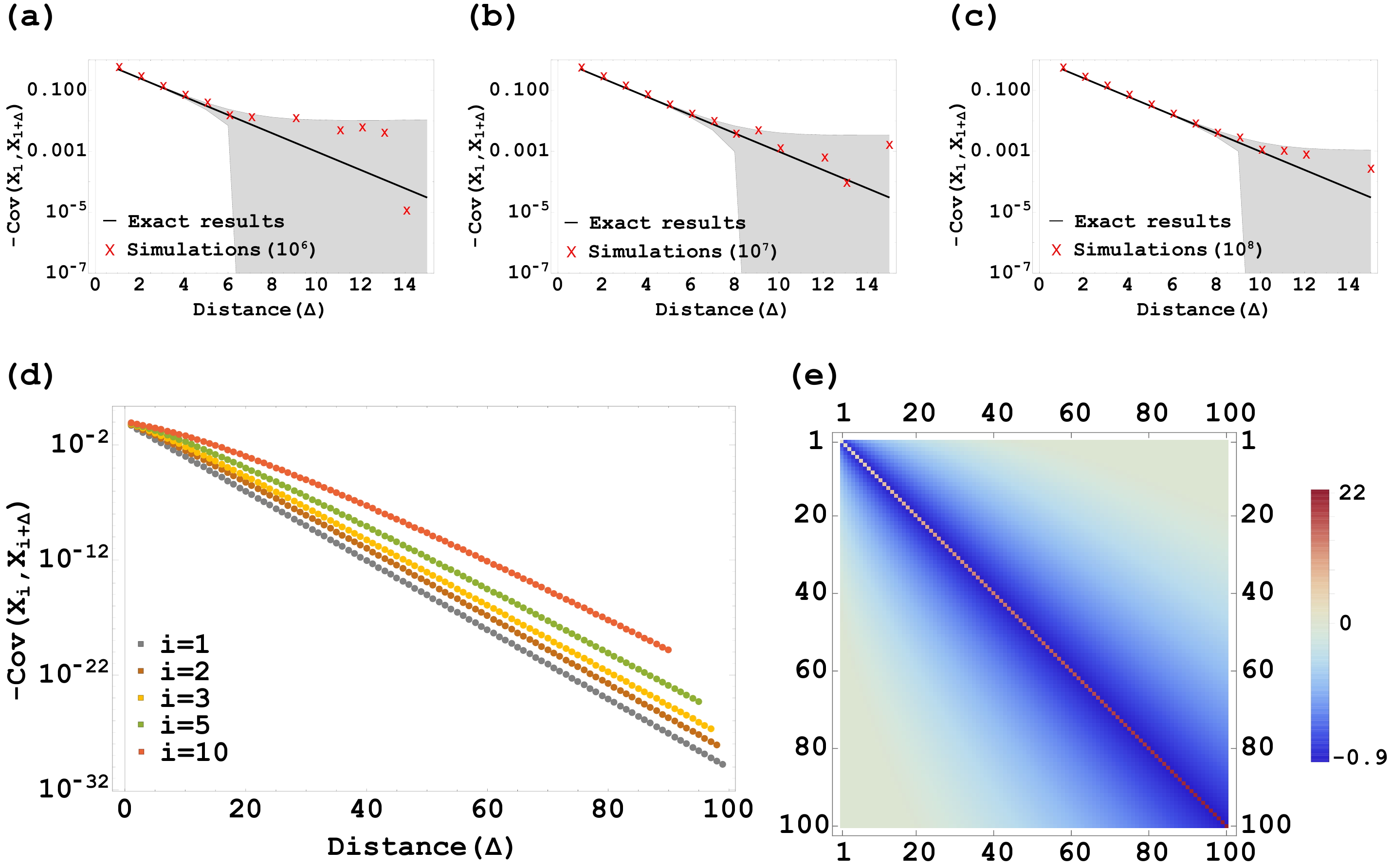}
\caption{Numerical validation and graphical illustration of the result in Eq. (\ref{homogeneous correlations}). Here, $n=100$ and $\lambda=\mu_1=\cdots=\mu_n=1$. Panels (a)-(c): comparison with numerical simulations. The exact result for the covariance between $X_1$ and $X_{1+\Delta}$ is plotted vs. the distance $\Delta$ between the two sites (solid black line). Note that the covariance is always negative and that it quickly drops with the distance. Results from simulations are presented as red crosses. These were obtained by sampling $10^6$, $10^7$ and $10^8$ occupancy vectors (panels a, b and c respectively). One can observe that the uncertainty around the numerical estimates (shaded in gray) slowly diminishes as the number of samples increases, thus converging to the black theory line. Panel (d): The exact result for the covariance between $X_i$ and $X_{i+\Delta}$ is plotted as a function of the distance $\Delta$ for different values of $i=1,2,3,5,10$. As expected correlations generally become weaker with distance, but now one can also observe that the decay is exponential. Panel (e): Heat map representation of the covariance matrix $\left\langle\mathbf{X}(n)\mathbf{X}(n)^{\top}\right \rangle$. Correlations are color coded: deep red/blue represent strong positive/negative correlation and white represents weak correlation.}
\label{Fig3}
\end{figure}
\newpage
We now turn to describe occupancy  correlations in homogeneous ASIPs for which all gate opening rates are equal: $\mu=\mu_1=\cdots=\mu_n$. In Sec. \textcolor{blue}{III}, we show that in this case the result in Eq. (\ref{general correlations}) can be simplified to give 
\begin{equation}
\left(\mathbf{Cov}\left[\mathbf{X}(n)\right]\right)_{i,j}=\begin{cases}
\left(\frac{1}{2^{i+j-1}}\left(\begin{array}{c}
j+i-1\\
\text{min}(i,j)
\end{array}\right){}_{2}F_{1}\left(1,i+j,1+\text{min}(i,j);\frac{1}{2}\right)-2\right)\rho^{2} & i\neq j\\
\\
\rho+\left(\frac{4\Gamma(i+1/2)}{\sqrt{\pi}\Gamma(i)}-1\right)\rho^{2} & i=j
\end{cases}\label{homogeneous correlations}
\end{equation}
where  ${\color{black}_{2}F_{1}\left(a,b,c;x\right)}$ is the hypergeometric function, \textcolor{black}{$\Gamma(x)$} is the Gamma function, and ${\color{black}\rho=\lambda/\mu}$ is the mean site occupancy. An interesting property of Eq. (\ref{homogeneous correlations}) is its invariance with respect to a change of time scale: the covariance matrix of a homogeneous ASIP with rates $\lambda$ and $\mu$ is identical to the covariance matrix of a homogeneous ASIP with rates $\alpha\lambda$ and $\alpha\mu$ ($\alpha>0$). In Sec. \textcolor{blue}{III}, we show that this is also true for inhomogeneous ASIP systems. The derivation of \eref{homogeneous correlations} is carried out in Sec. \textcolor{blue}{III}. For numerical validation and graphical illustration of this result see Fig. \ref{Fig3}.

\newpage
\section{CORRELATION MATRIX: EXACT RESULTS}

\noindent In this section, we derive and solve an equation for the cross-correlations matrix of an ASIP in steady-state.

\begin{center}
\textbf{\textcolor{black}{A. Temporal dynamics of the cross-correlations matrix}}
\par\end{center}

We start by deriving an equation for the temporal dynamics of the cross-correlations matrix \textcolor{black}{$\left\langle \mathbf{X}(n,t)\mathbf{X}^{\top}(n,t)\right\rangle$}. Consider an ASIP of $n$ sites and observe the system at times $t$ and ${\color{black}t'=t+\Delta}$ (for small $\Delta$). We would like to find a relation between the cross-correlations matrices \textcolor{black}{$\left\langle \mathbf{X}(n,t^{\prime})\mathbf{X}^{\top}(n,t^{\prime})\right\rangle $} and \textcolor{black}{$\left\langle \mathbf{X}(n,t)\mathbf{X}^{\top}(n,t)\right\rangle $}. To do so, we utilize the Markovian law of motion in Eq. (\ref{markov dynamics}) to compute $\mathbf{X}(n,t')$ given \textcolor{black}{$ \mathbf{X}(n,t)$}. We find 
\begin{equation}
    \mathbf{X}(n,t')|\mathbf{X}(n,t)=\begin{cases}
\mathbf{X}(n,t)   &  \text{no arrival or gate opening} \\
\mathbf{X}(n,t)+\mathbf{\mathbf{e_{1,n}}} & \text{arrival to the 1\textsuperscript{st} site}\\
\mathbf{X}(n,t)+\mathbf{\mathbf{M_{k,n}}}\mathbf{X}(n,t) & \text{opening of the k\textsuperscript{th} gate}\:,
\end{cases}
\end{equation}
where the vector $\mathbf{e_{1,n}}$ and the operators $\mathbf{\mathbf{M_{k,n}}}$ where defined in Eqs. (\ref{1 vector}) and (\ref{Mk operator}) respectively.  
Now, we utilize the law of total expectation to obtain
\begin{equation}
{\begin{array}{l}
\left\langle \mathbf{X}(n,t^{\prime})\mathbf{X}^{\top}(n,t^{\prime})\right\rangle =\left\{ \begin{array}{l}
\left(1-\left(\lambda+\mu_{c}\right)\Delta\right)\left\langle \mathbf{X}(n,t)\mathbf{X}^{\top}(n,t)\right\rangle \\
+\\
\lambda\Delta\left\langle (\mathbf{X}(n,t)+\mathbf{e_{1,n}})(\mathbf{X}(n,t)+\mathbf{e_{1,n}})^{\top}\right\rangle \\
+\\
\mu_{1}\Delta\left\langle (\mathbf{X}(n,t)+\mathbf{\mathbf{M_{1,n}}}\mathbf{X}(n,t))(\mathbf{X}(n,t)+\mathbf{\mathbf{M_{1,n}}}\mathbf{X}(n,t))^{\top}\right\rangle \\
+\\
\mu_{2}\Delta\left\langle (\mathbf{X}(n,t)+\mathbf{\mathbf{M_{2,n}}}\mathbf{X}(n,t))(\mathbf{X}(n,t)+\mathbf{\mathbf{M_{2,n}}}\mathbf{X}(n,t))^{\top}\right\rangle \\
+\cdots+\\
\mu_{\mathrm{n-1}}\Delta\left\langle (\mathbf{X}(n,t)+\mathbf{\mathbf{M_{n-1,n}}}\mathbf{X}(n,t))(\mathbf{X}(n,t)+\mathbf{\mathbf{M_{n-1,n}}}\mathbf{X}(n,t))^{\top}\right\rangle \\
+\\
\mu_{\mathrm{n}}\Delta\left\langle (\mathbf{X}(n,t)+\mathbf{\mathbf{M_{n,n}}}\mathbf{X}(n,t))(\mathbf{X}(n,t)+\mathbf{\mathbf{M_{n,n}}}\mathbf{X}(n,t))^{\top}\right\rangle\\ 
+\\
o(\Delta)\,\,,
\end{array}\right.
\end{array}}\label{cross correlation}
\end{equation}
where we recall that $\mu_{c}=\mu_{1}+\cdots+\mu_{n}$. 

Rearranging terms in Eq. (\ref{cross correlation}),
dividing by $\Delta$, and taking $\Delta\rightarrow0$, we find that 
\begin{equation}
{\color{black}\frac{d\left\langle \mathbf{X}(n,t)\mathbf{X}^{\top}(n,t)\right\rangle }{dt}=\left\{ \begin{array}{l}
\lambda\left[\left\langle \mathbf{X}(n,t)\right\rangle \mathbf{e_{1,n}^{\top}}+\mathbf{e_{1,n}}\left\langle \mathbf{X}(n,t)^{\top}\right\rangle +\mathbf{e_{1,n}}\mathbf{e_{1,n}^{\top}}\right]\\
+\\
\left\langle \mathbf{X}(n,t)\mathbf{\mathbf{X}^{\top}}(n,t)\right\rangle \left(\overset{n}{\underset{i=1}{\sum}}\mu_{\mathrm{i}}\mathbf{\mathbf{M_{i,n}}}\right)^{\top}\\
+\\
\left(\overset{n}{\underset{i=1}{\sum}}\mathrm{\mu_{i}}\mathbf{M_{i,n}}\right)\left\langle \mathbf{X}(n,t)\mathbf{\mathbf{X}^{\top}}(n,t)\right\rangle \\
+\\
\overset{n}{\underset{i=1}{\sum}}\mathrm{\mu_{i}}\mathbf{M_{i,n}}\left\langle \mathbf{\mathbf{X}}(n,t)\mathbf{X}^{\top}(n,t)\right\rangle \mathbf{M_{i,n}^{\top}}.
\end{array}\right.}\label{ode cross correlation}
\end{equation}
\noindent Equation (\ref{ode cross correlation}) governs the temporal dynamics of the cross-correlations matrix \textcolor{black}{$\left\langle \mathbf{X}(n,t)\mathbf{X}^{\top}(n,t)\right\rangle $} in the ASIP. In what follows, we will also specifically require $\left\langle \mathrm{X}_{\mathrm{n}}(t)\mathbf{X}^{\top}(n,t)\right\rangle$, i.e., the last row of the cross-correlations matrix $\left\langle \mathbf{X}(n,t)\mathbf{X}^{\top}(n,t)\right\rangle$. Equation (\ref{ode cross correlation}) can be used to write an equation for $\left\langle \mathrm{X}_{\mathrm{n}}(t)\mathbf{X}^{\top}(n,t)\right\rangle$, but it is actually easier to derive this equation from scratch.

Consider the same ASIP system as before and observe it at times $t$ and ${\color{black}t'=t+\Delta}$ (for small $\Delta$). Following the same recipe that led to Eq. (\ref{cross correlation}), we find that the relation between  $\left\langle\mathrm{X}_{\mathrm{n}}(t^{\prime})\mathbf{X}^{\top}(n,t^{\prime})\right\rangle$ and $\left\langle \mathrm{X}_{\mathrm{n}}(t)\mathbf{X}^{\top}(n,t)\right\rangle$ is given by

\begin{equation}
\begin{array}{l}
\left\langle \mathrm{X}_{\mathrm{n}}(t^{\prime})\mathbf{X}^{\top}(n,t^{\prime})\right\rangle =\left\{ \begin{array}{l}
\left(1-\left(\lambda+\mu_{c}\right)\Delta\right)\left\langle \mathrm{X}_{\mathrm{n}}(t)\mathbf{X}^{\top}(n,t)\right\rangle \\
+\\
\lambda\Delta\left\langle \mathrm{X}_{\mathrm{n}}(t)(\mathbf{X}(n,t)+\mathbf{e_{1,n}})^{\top}\right\rangle \\
+\\
\mu_{1}\Delta\left\langle \mathrm{X}_{\mathrm{n}}(t)(\mathbf{X}(n,t)+\mathbf{\mathbf{M_{1,n}X}}(n,t))^{\top}\right\rangle \\
+\\
\mu_{2}\Delta\left\langle \mathrm{X}_{\mathrm{n}}(t)(\mathbf{X}(n,t)+\mathbf{\mathbf{M_{2,n}X}}(n,t))^{\top}\right\rangle \\
+\cdots+\\
\mu_{n-1}\Delta\left\langle (\mathrm{X}_{\mathrm{n}}(t)+\mathrm{X}_{\mathrm{n-1}}(t))(\mathbf{X}(n,t)+\mathbf{\mathbf{M_{n-1,n}X}}(n,t))^{\top}\right\rangle \\
+\\
\mu_{n}\Delta\left\langle 0\cdot(\mathbf{X}(n,t)+\mathbf{\mathbf{M_{n,n}X}}(n,t))^{\top}\right\rangle \\
+\\
o(\Delta)\,\,.
\end{array}\right.\text{ }\end{array}\label{cross correlation row}
\end{equation}

\noindent Rearranging terms in Eq. (\ref{cross correlation row}),
dividing by \textcolor{black}{$\Delta$, }and taking $\Delta\rightarrow0$, we find

\begin{equation}
{\color{black}\frac{d\left\langle \mathrm{X}_{\mathrm{n}}(t)\mathbf{X}^{\top}(n,t)\right\rangle }{dt}=\left\{ \begin{array}{l}
\lambda\left\langle \mathrm{X}_{\mathrm{n}}(t)\right\rangle \mathbf{e_{1,n}^{\top}}\\
+\\
\left\langle \mathrm{\mathrm{X}_{n}(}t)\mathbf{\mathbf{X}^{\top}}(n,t)\right\rangle \left[\overset{n-1}{\underset{i=1}{\sum}}\mu_{\mathrm{i}}\mathbf{\mathbf{M_{i,n}^{\top}}}-\mu_{\mathrm{n}}\mathbf{I}\right]\\
+\\
\mu_{\mathrm{n-1}}\left\langle \mathrm{X_{n-1}}\mathbf{\mathbf{X}^{\top}}(n,t)\right\rangle \left[\mathbf{\mathbf{M_{n-1,n}^{\top}}}+\mathbf{I}\right].
\end{array}\right.}\label{ode cross correlation row}
\end{equation}
\\

\begin{center}
\textbf{\textcolor{black}{B. Steady-state}}
\par\end{center}

In this subsection, we derive an explicit steady-state solution for Eq. (\ref{ode cross correlation}). We do so in several steps. In the first step, we establish a recursion relation by finding the steady-state solution for $\left\langle \mathrm{X}_{\mathrm{n}}(t)\mathbf{X}^{\top}(n,t)\right\rangle$ in Eq. (\ref{ode cross correlation row}) and showing that it can be written down in terms  of the steady-state solution for $\left\langle \mathrm{X}_{\mathrm{n-1}}(t)\mathbf{X}^{\top}(n-1,t)\right\rangle$, i.e., the last row of the cross-correlations matrix of an ASIP that lacks the $n\textsuperscript{th}$ site. We start by noting that in steady-state Eq. (\ref{ode cross correlation row}) reduces to
\begin{equation}
\begin{array}{l}
0={\color{black}\left\{ \begin{array}{l}
\lambda^{2}\mathbf{e_{1,n}^{\top}}/\mu_{\mathrm{n}}\\
+\\
\left\langle \mathrm{X_{n}}\mathbf{\mathbf{X}^{\top}}(n)\right\rangle \left[\overset{n-1}{\underset{i=1}{\sum}}\mu_{\mathrm{i}}\mathbf{\mathbf{M_{i,n}^{\top}}}-\mu_{\mathrm{n}}\mathbf{I}\right]\\
+\\
\mu_{\mathrm{n-1}}\left\langle \mathrm{X_{n-1}}\mathbf{\mathbf{X}^{\top}}(n)\right\rangle \left[\mathbf{\mathbf{M_{n-1,n}^{\top}}}+\mathbf{I}\right],
\end{array}\right.}
\end{array}\label{steady state equations cross correlations}
\end{equation}
\noindent where we have used $\left\langle \mathbf{\mathrm{X}_{\mathrm{n}}}\right\rangle =\lambda/\mu_{n}$ [Eq. (\ref{mean occupation steady state})]. In order to proceed, we  will simplify $\left\langle \mathrm{X_{n-1}}\mathbf{\mathbf{X}^{\top}}(n)\right\rangle$ that appears in the bottom row of \eref{steady state equations cross correlations}. To this end, we will make use of the ``embedding" property of the ASIP \cite{ASIP-1}. 

The embedding property is described as follows. Consider two ASIP systems: (A) and (B), and set the number of sites in these systems to $n_A$ and $n_B$ respectively where $n_A\leq n_B$. Assume that the inflow rate $\lambda$, and the opening rates of the first $n_A$ gates, are identical in systems A and B. It can then be shown that the steady-state occupancy of model system (A) is statistically identical to the steady-state occupancy of the first $n_A$ sites of model system (B). In other words, system (A) will have the same steady-state when it stands alone and when it is embedded within system (B) [as long as the embedding is done in the manner described above]. To intuitively understand the embedding property one simply needs to realize that a part of an ASIP system cannot feed-back on a disjoint part of the system which is located upstream with respect to it (closer to the first site). Thus, the steady-state distribution of the first $n_A$ sites in system (B) cannot depend on the remaining $n_B-n_A$ sites that are located downstream to it. A formal proof of the embedding property was given in \cite{ASIP-1}.  

The embedding property of the ASIP implies that the steady-state occupancy distribution of the first $n-1$ sites in an ASIP of $n$ sites is identical to the steady-state occupancy distribution in an ASIP of $n-1$ sites that has the same inflow and gate opening rates. Using this, we can write $\left\langle \mathrm{X}_{\mathrm{n-1}}\mathbf{\mathbf{X}^{\top}}(n)\right\rangle$ which appears on the right hand side of \eref{steady state equations cross correlations} as

\begin{equation}
\left\langle \mathrm{X}_{\mathrm{n-1}}\mathbf{\mathbf{X}^{\top}}(n)\right\rangle =\left(\left\langle \mathrm{X}_{\mathrm{n-1}}\mathbf{\mathbf{X}^{\top}}(n-1)\right\rangle ,0\right)+\left( 0,\cdots,0,\langle\mathrm{X}_{\mathrm{n-1}}\mathrm{X}_{\mathrm{n}}\rangle\right)\,\,,
\end{equation}
and further note that
\begin{equation}
\left\langle \mathrm{X}_{\mathrm{n-1}}\mathbf{\mathbf{X}^{\top}}(n)\right\rangle =\left(\left\langle \mathrm{X}_{\mathrm{n-1}}\mathbf{\mathbf{X}^{\top}}(n-1)\right\rangle ,0\right)+\left(\mathbf{P_{n}}\left\langle \mathrm{X}_{\mathrm{n}}\mathbf{\mathbf{X}}(n)\right\rangle \right)^{\top}\,\,,\label{P_embedding}
\end{equation}
where we have used the operator {$\mathbf{P_{n}}$} that was defined in Eq. (\ref{P operator}). Substituting Eq. (\ref{P_embedding}) into Eq. (\ref{steady state equations cross correlations}), noting that
\begin{equation}
\left(\mathbf{P_{n}}\left\langle \mathrm{X_{n}}\mathbf{\mathbf{X}}(n)\right\rangle \right)^{\top}\mathbf{\mathbf{M_{n-1,n}^{\top}}}=\left(\mathbf{\mathbf{M_{n-1,n}}}\left(0,\cdots,0,\langle\mathrm{X}_{\mathrm{n-1}}\mathrm{X}_{\mathrm{n}}\rangle\right)\right)^{\top}=\mathbf{0}\,,
\end{equation}
and rearranging, yields the following recursion relation 
\begin{equation}
\left\langle \mathrm{X}_{\mathrm{n}}\mathbf{\mathbf{X}^{\top}}(n)\right\rangle =-\left[\frac{\lambda^{2}}{\mu_{\mathrm{n}}}\mathbf{e_{1,n}^{\top}}+\mu_{\mathrm{n-1}}\left(\left\langle \mathrm{X}_{\mathrm{n-1}}\mathbf{\mathbf{X^{\top}}}(n-1)\right\rangle ,0\right)\left[\mathbf{\mathbf{M_{n-1,n}^{\top}}}+\mathbf{I}\right]\right]\mathrm{\mathbf{R}_{\mathbf{n}}^{-1}},\label{recursion}
\end{equation}
where the matrix ${\color{black}\mathrm{\mathbf{R}_{\mathbf{n}}}}$ was defined in Eq. (\ref{R matrix}). Equation (\ref{recursion}) connects $\left\langle \mathrm{X}_{\mathrm{n}}\mathbf{X}^{\top}(n)\right\rangle$ with $\left\langle
\mathrm{X}_{\mathrm{n-1}}\mathbf{\mathbf{X}^{\top}}(n-1)\right\rangle$ and thus provides an iterative method for computing the
cross-correlations matrix. 

For $n>2$, \eref{recursion} can be further simplified. In order to do so, we utilize the embedding property one more time to obtain 
\begin{equation}
\left\langle \mathrm{X}_{\mathrm{n}}\mathbf{\mathbf{X}^{\top}}(n)\right\rangle =-\left[\frac{\lambda^{2}}{\mu_{\mathrm{n}}}\mathbf{e_{1,n}^{\top}}+\mu_{\mathrm{n-1}}\left(\left\langle \mathrm{X}_{\mathrm{n-1}}\mathbf{\mathbf{X^{\top}}}(n-2)\right\rangle ,\left\langle \mathrm{X}_{\mathrm{n-1}}\mathrm{\mathit{\mathrm{X}}_{n-1}}\right\rangle ,0\right)\left[\mathbf{\mathbf{M_{n-1,n}^{\top}}}+\mathbf{I}\right]\right]\mathbf{\mathbf{R}_{n}^{-1}}\:.\label{recursion 2}
\end{equation}
Noting that (for $n\geq 2$)

\begin{equation}
\mathbf{\mathbf{M_{n-1,n}^{\top}}}+\mathbf{I}=\left(\begin{array}{ccccc}
1 &  &  & {\color{black}0}\\
 & \ddots &  & {\color{black}0}\\
 &  & 1 & {\color{black}\vdots}\\
 &  & \cdots & 0 & 1\\
 &  & \cdots & {\color{black}0} & 1
\end{array}\right),
\end{equation}
and taking the vector-matrix product inside the parenthesis of \eref{recursion 2} we find

\begin{equation}
{\color{black}\left\langle \mathrm{X}_{\mathrm{n}}\mathbf{\mathbf{X}^{\top}}(\mathit{n})\right\rangle =-\left[\frac{\lambda^{2}}{\mu_{n}}{\mathbf{e_{1,n}^{\top}}}+\mu_{n-1}\left(\left\langle \mathrm{X}_{\mathrm{n-1}}\mathbf{\mathbf{X}^{\top}}(\mathit{n\mathrm{-2}})\right\rangle ,0,\left\langle \mathrm{X}_{\mathrm{n-1}}\mathrm{X}_{\mathrm{n-1}}\right\rangle \right)\right]\mathbf{\mathbf{R}_{n}^{-1}}}.\label{recursion 3}
\end{equation}
Finally, noting that  
\begin{align}
{\mu_{n-1}\left(\left\langle \mathrm{X}_{\mathrm{n-1}}\mathbf{\mathbf{X}}(n-2)\right\rangle ^{\mathrm{\top}},0,\left\langle \mathrm{X}_{\mathrm{n-1}}\mathrm{X}_{\mathrm{n-1}}\right\rangle \right)=\left\langle \mathrm{X}_{\mathrm{n-1}}\mathbf{\mathbf{X}}(n-1)^{\top}\right\rangle \mathbf{\mathrm{\mathbf{Q}}_{\mathrm{\mathbf{n}}}}}\:,\label{Q matrix def}
\end{align}
where $\mathbf{\mathrm{\mathbf{Q}}_{\mathrm{\mathbf{n}}}}$ was defined in \eref{Q matrices}, we conclude that
\begin{align}
\left\langle \mathrm{X}_{\mathrm{n}}\mathbf{\mathbf{X}^{\top}}(n)\right\rangle =-\mathrm{\mathbf{V}}_{\mathrm{\mathbf{n}}}\mathbf{\mathbf{R}_{n}^{-1}}-{\color{blue}{\color{black}\left\langle \mathrm{X}_{\mathrm{n-1}}\mathbf{\mathbf{X^{\top}}}(n-1)\right\rangle \mathbf{Q}_{\mathbf{n}}\mathbf{\mathbf{R}_{n}^{-1}}}}\:,\label{recursion 4}
\end{align}
where $\mathrm{\mathbf{V}}_{\mathrm{n}}$ was defined in \eref{V vector}. Applying the iteration in \eref{recursion 4} time and again one can solve for the cross correlation matrix of an arbitrary ASIP system of size $n > 2$. The base for these iterations (seed) is the cross correlation matrix of an ASIP system of size $n=2$ which we now compute. The cross correlation matrix of an ASIP system of size $n=1$ will also be obtained as a preliminary step. 

In order to obtain a steady-state solution for an ASIP system of size $n=1$ we substitute Eqs. (\ref{1 vector}), (\ref{Mk operator}) and (\ref{Mn operator}) into Eq. (\ref{ode cross correlation}) and set the left-hand side of this equation to zero. This gives

\begin{equation}
    0=\lambda\left(2\left\langle \mathrm{X}_{1}\right\rangle +1\right)-\mu_{1}\left\langle \mathrm{X}_{1}^{2}\right\rangle -\mu_{1}\left\langle \mathrm{X}_{1}^{2}\right\rangle +\mu_{1}\left\langle \mathrm{X}_{1}^{2}\right\rangle . \label{seed 1}
\end{equation}
Equation (\ref{seed 1}) can be solved by substituting $\left\langle \mathbf{\mathrm{X}_{\mathrm{1}}}\right\rangle =\lambda/\mu_{1}$ [Eq. (\ref{mean occupation steady state})] and rearranging terms to give

\begin{equation}
    \left\langle \mathrm{X_{1}^{2}}\right\rangle =\frac{\lambda(2\lambda+\mu_{1})}{\mu_{1}^{2}}\:.\label{seed 2}
\end{equation}

The solution for $n=2$, $\left\langle \mathrm{X_{2}}\mathbf{\mathbf{X}^{\top}}(2)\right\rangle$, can be obtained by solving \eref{recursion} for $n=2$. To do so, we start by calculating $\mathbf{\mathbf{R}_{2}}$ [\eref{R matrix}]
\begin{equation}
    \mathbf{\mathbf{\mathbf{R}_{2}}=\mathbf{\mathrm{\mu_{1}}\mathbf{\mathbf{M_{1,2}^{\top}}}+\mathrm{\mu_{1}}\mathbf{P_{2}^{\top}}-\mathrm{\mu_{2}}\mathbf{I}=\left(\begin{array}{cc}
-\mu_{1}-\mu_{2} & 2\mu_{1}\\
0 & -\mu_{2}
\end{array}\right)}}. \label{R2}
\end{equation}
Now, we observe that $\mathbf{\mathbf{R}_{2}^{-1}}$ is given by
\begin{equation}
    \mathbf{\mathbf{R}_{2}^{-1}}=\left(\begin{array}{cc}
\frac{-1}{\mu_{1}+\mu_{2}} & \frac{-2\mu_{1}}{\mu_{2}(\mu_{1}+\mu_{2})}\\
0 & \frac{-1}{\mu_{2}}
\end{array}\right). \label{R2 inv}
\end{equation}
Substituting \eref{1 vector}, (\ref{Mk operator}) and (\ref{R2 inv}) into Eq. (\ref{recursion}) then yields
\begin{equation}
    \left\langle \mathrm{X}_{2}\mathbf{\mathbf{X}^{\top}}(2)\right\rangle =-\left[\frac{\lambda^{2}}{\mu_{2}}\left(\begin{array}{cc}
1, & 0\end{array}\right)+\mu_{1}\left(\begin{array}{cc}
\left\langle \mathrm{X_{1}X_{1}}\right\rangle,  & 0\end{array}\right)\left[\left(\begin{array}{cc}
-1 & 1\\
0 & 0
\end{array}\right)+\left(\begin{array}{cc}
1 & 0\\
0 & 1
\end{array}\right)\right]\right]\left(\begin{array}{cc}
\frac{-1}{\mu_{1}+\mu_{2}} & \frac{-2\mu_{1}}{\mu_{2}(\mu_{1}+\mu_{2})}\\
0 & -\frac{1}{\mu_{2}}
\end{array}\right).\label{n=2 eq}
\end{equation}
Substituting \eref{seed 2} into \eref{n=2 eq} we find
\begin{equation}
    \left\langle \mathrm{X}_{2}\mathbf{\mathbf{X}^{\top}}(2)\right\rangle =\left(\begin{array}{cc}
\frac{\lambda^{2}}{\mu_{2}(\mu_{1}+\mu_{2})}\:, & \frac{2\mu_{1}\lambda^{2}}{\mu_{2}^{2}(\mu_{1}+\mu_{2})}+\frac{\lambda(2\lambda+\mu_{1})}{\mu_{1}\mu_{2}}\end{array}\right), \label{solution n=2}
\end{equation}
Which can be brought into the form of \eref{recursion 4} by noting that 
\begin{equation}
    \left\langle \mathrm{X}_{\mathrm{2}}\mathbf{\mathbf{X}^{\top}}(2)\right\rangle =-\mathbf{V}_{\mathbf{2}}\mathbf{\mathbf{R}_{2}^{-1}}-\frac{\lambda(2\lambda+\mu_{1})}{\mu_{1}^{2}}\mathbf{Q}_{\mathbf{2}}\mathbf{\mathbf{R}_{2}^{-1}}, \label{itr n=2}
\end{equation}
with $\mathbf{Q}_{\mathbf{2}}=\left(\begin{array}{cc}
0, & \mu_{1}\end{array}\right)$.  

Solutions for ASIP systems with $n>2$ can now be obtained in terms of $\mathbf{V}_{\mathbf{n}}, \mathbf{\mathbf{R}_{n}}$ and $\mathbf{\mathbf{Q}_{n}}$ by substituting \eref{itr n=2} into \eref{recursion 4} and progressing iteratively. Doing this, we find 
\begin{equation}
\begin{gathered}\left\langle \mathrm{X}_{\mathrm{3}}\mathbf{\mathbf{X}^{\top}}(3)\right\rangle =-\mathbf{V}_{\mathbf{3}}\mathbf{\mathbf{R}_{3}^{-1}}+\mathbf{V}_{\mathbf{2}}\mathbf{\mathbf{R}_{2}^{-1}\mathbf{Q}_{\mathbf{3}}\mathbf{\mathbf{R}_{3}^{-1}}}+\frac{\lambda(2\lambda+\mu_{1})}{\mu_{1}^{2}}\mathbf{Q}_{\mathbf{2}}\mathbf{\mathbf{R}_{2}^{-1}\mathbf{Q}_{\mathbf{3}}\mathbf{\mathbf{R}_{3}^{-1}}}\\
\left\langle \mathrm{X}_{\mathrm{4}}\mathbf{\mathbf{X}^{\top}}(4)\right\rangle =-\mathbf{V}_{\mathbf{4}}\mathbf{\mathbf{R}_{4}^{-1}}+\mathbf{V}_{\mathbf{3}}\mathbf{\mathbf{R}_{3}^{-1}}\mathbf{Q}_{\mathbf{4}}\mathbf{\mathbf{R}_{4}^{-1}}-\mathbf{V}_{\mathbf{2}}\mathbf{\mathbf{R}_{2}^{-1}\mathbf{Q}_{\mathbf{3}}\mathbf{\mathbf{R}_{3}^{-1}}}\mathbf{Q}_{\mathbf{4}}\mathbf{\mathbf{R}_{4}^{-1}}-\frac{\lambda(2\lambda+\mu_{1})}{\mu_{1}^{2}}\mathbf{Q}_{\mathbf{2}}\mathbf{\mathbf{R}_{2}^{-1}\mathbf{Q}_{\mathbf{3}}\mathbf{\mathbf{R}_{3}^{-1}}}\mathbf{Q}_{\mathbf{4}}\mathbf{\mathbf{R}_{4}^{-1}}.
\end{gathered}
\label{cross correlations solutions}
\end{equation}
Continuing in a similar fashion, we obtain the general formula in \eref{general correlations}.

An interesting property of the covariance matrix of the ASIP in steady-state is its invariance with respect to a change of time scale. Consider an ASIP system (A) with ${\color{black}n}$ sites and parameters ${\color{black}\left\{ \lambda,\mu_{1},\mu_{2},\ldots,\mu_{n}\right\} }$ and an ASIP system (B) with $n$ sites and parameters $\left\{ \alpha\lambda,\alpha\mu_{1},\alpha\mu_{2},\ldots,\alpha\mu_{n}\right\}$, where $\alpha>0$ is a nonzero scalar. Substituting the rates of both systems into Eqs. (\ref{V vector}, \ref{R matrix}, \ref{Q matrices}) yields
\begin{equation}
\begin{cases}
\mathbf{R_{n}^{\mathit{\boldsymbol{B}}}\mathbf{\mathrm{=\mathit{\alpha}}R_{n}^{\mathit{\boldsymbol{A}}}}}\\
\mathbf{\mathrm{(}R_{\mathbf{n}}^{\mathit{\boldsymbol{B}}}\mathrm{)}^{\mathrm{-1}}}=\mathbf{\mathbf{\frac{1}{\mathit{\alpha}}}\mathrm{(}R_{\mathbf{n}}^{\mathit{\boldsymbol{A}}}\mathrm{)}^{\mathrm{-1}}}\\
\mathbf{Q}_{\mathbf{n}}^{\boldsymbol{B}}\mathit{=\alpha}\mathbf{Q}_{\mathbf{n}}^{\boldsymbol{A}}\\
\mathrm{\mathbf{V}}_{\mathrm{\mathbf{n}}}^{\boldsymbol{B}}=\mathit{\alpha}\mathrm{\mathbf{V}}_{\mathrm{\mathbf{n}}}^{\boldsymbol{A}}
\end{cases}\label{operators scaling}
\end{equation}
Substituting Eq. (\ref{operators scaling}) into Eq. (\ref{general correlations}) and applying \textcolor{black}{the algorithm illustrated in Fig. }\textcolor{blue}{1}, we obtain that the cross-correlations matrices of systems (A) and (B) are identical. In addition, it is easy to see that Eq. (\ref{means matrix}) yields the same result for both systems. These results, together with Eq. (\ref{covariance steady state}), prove invariance with respect to a change of time scale.

\section{CORRELATION MATRIX: HOMOGENEOUS SYSTEMS}

\noindent In Sec. \textcolor{blue}{III}, we presented results for general ASIPs. In this section, we restrict ourselves to the case of homogeneous ASIPs, i.e., systems in which all gate opening rates are identical. For these ASIPs, we present a simple formula for the covariance matrix and discuss its special properties. We start by writing explicit expressions for the vector-matrix and matrix-matrix products which appear in  \eref{general correlations}. 

Consider a homogeneous ASIP where $\mu=\mu_{\mathbf{\mathrm{1}}}=\ldots=\mu_{\mathrm{n}}$. Equation (\ref{R matrix explicit form}) then reduces to 
\begin{equation}
{\color{black}\mathbf{\mathbf{R}_{\mathrm{\mathbf{n}}}}=\left(\begin{array}{ccccc}
-2\mu & \mu\\
 & -2\mu & \ddots\\
 &  & \ddots & \mu\\
 &  &  & -2\mu & 2\mu\\
 &  &  &  & -\mu
\end{array}\right)},\label{homogeneous R}
\end{equation}
and the inverse matrix, $\mathrm{\mathbf{\mathbf{R}_{n}^{-1}}}$, is then given by
\begin{equation}
\mathbf{\mathbf{R}_{\mathrm{\mathrm{\mathbf{n}}}}^{-1}}=\left(\begin{array}{cccccc}
-\frac{1}{2\mu} & -\frac{1}{4\mu} & \cdots & -\frac{1}{2^{n-2}\mu} & -\frac{1}{2^{n-1}\mu} & -\frac{1}{2^{n-2}\mu}\\
 & -\frac{1}{2\mu} & \cdots & -\frac{1}{2^{n-3}\mu} & -\frac{1}{2^{n-2}\mu} & -\frac{1}{2^{n-3}\mu}\\
 &  & \ddots & \vdots & \vdots & \vdots\\
 &  &  &  & -\frac{1}{4\mu} & -\frac{1}{2\mu}\\
 &  &  &  & -\frac{1}{2\mu} & -\frac{1}{\mu}\\
 &  &  &  &  & -\frac{1}{\mu}
\end{array}\right).\label{homogeneous R-1}
\end{equation}
Now, using the definition of $\mathrm{\mathbf{V}}_{\mathbf{\mathrm{\mathbf{i}}}}$ in \eref{V vector}, we compute ${\color{black}\mathrm{\mathbf{\mathrm{\mathbf{V}}_{\mathrm{\mathrm{\mathbf{i}}}}\mathbf{\mathbf{R}_{i}^{-1}}}}}$ and find \begin{equation}
\mathrm{\mathbf{V}}_{\mathbf{\mathrm{\mathbf{2}}}}\mathbf{\mathbf{R}_{2}^{-1}}=-\left(\frac{\lambda}{\mu}\right)^{2}\left(\begin{array}{cc}
\frac{1}{2}, & 1\end{array}\right),
\end{equation}
for $i=2$, and 
\begin{equation}
\mathrm{\mathbf{V}}_{\mathbf{\mathrm{\mathbf{i}}}}\mathbf{\mathbf{R}_{i}^{-1}}=-\left(\frac{\lambda}{\mu}\right)^{2}\left(\begin{array}{ccccc}
\frac{1}{2}, & \frac{1}{4}, & \cdots, & \mathrm{\frac{1}{2^{i-1}}}, & \mathrm{\frac{1}{2^{i-2}}}\end{array}\right),\label{VR-1}
\end{equation}
for $i>2$. Similarly, using the definition of $\mathrm{\mathbf{Q}_{\mathrm{\mathbf{i}}}}$ given in \eref{Q matrices} and above it, we compute $\mathrm{\mathbf{Q}_{\mathbf{i}}}\mathbf{\mathbf{R}_{i}^{-1}}$ and find 
\begin{equation}
    \mathrm{\mathbf{Q}_{\mathbf{2}}}\mathbf{\mathbf{R}_{2}^{-1}}=\left(\begin{array}{cc}
0, & -1\end{array}\right)
\end{equation} for $i=2$, and
\begin{equation}
\mathrm{\mathbf{Q}_{\mathbf{i}}}\mathbf{\mathbf{R}_{i}^{-1}}=\left(\mathrm{\begin{array}{cccccc}
-\frac{1}{2} & -\frac{1}{4} & \cdots & -\frac{1}{2^{n-2}} & -\frac{1}{2^{n-1}} & -\frac{1}{2^{n-2}}\\
0 & -\frac{1}{2} & \cdots & \frac{1}{2^{n-3}} & -\frac{1}{2^{n-2}} & -\frac{1}{2^{n-3}}\\
0 & 0 & \ddots & \vdots & \vdots & \vdots\\
0 & 0 & 0 & -\frac{1}{2} & -\frac{1}{4} & -\frac{1}{2}\\
0 & 0 & 0 & 0 & 0 & -1
\end{array}}\right)\label{QR-1}
\end{equation}
for $i>2$. 

Evaluating the cross-correlations matrix $\left\langle\mathbf{X}(n)\mathbf{X}(n)^{\top}\right \rangle$ of a homogeneous ASIP system of arbitrary size can be done by substituting Eqs. (\ref{homogeneous R-1}-\ref{QR-1}) back into Eq. (\ref{general correlations}) and carrying out the computation using a standard computer algebra software. The covariance matrix then follows immediately and reveals a striking regularity. To demonstrate it, we consider a  homogeneous ASIP with six sites. Setting $\rho=\lambda/\mu$, we find that the covariance matrix for this system is given by
\bea
\lefteqn{\mathbf{Cov}\left[\mathbf{X}(6)\right]=}\nonumber\\
&&\left(\begin{array}{cccccc}
\rho^{2}+\rho\qquad & -\frac{1}{2}\rho^{2} & -\frac{1}{4}\rho^{2} & -\frac{1}{8}\rho^{2} & -\frac{1}{16}\rho^{2} & -\frac{1}{32}\rho^{2}\\
-\frac{1}{2}\rho^{2} & \qquad2\rho^{2}+\rho & -\frac{5}{8}\rho^{2} & -\frac{6}{16}\rho^{2} & -\frac{7}{32}\rho^{2} & -\frac{8}{64}\rho^{2}\\
-\frac{1}{4}\rho^{2} & -\frac{5}{8}\rho^{2} & \qquad\frac{11}{4}\rho^{2}+\rho\qquad & -\frac{22}{32}\rho^{2} & -\frac{29}{64}\rho^{2} & -\frac{37}{128}\rho^{2}\\
-\frac{1}{8}\rho^{2} & -\frac{6}{16}\rho^{2} & -\frac{22}{32}\rho^{2} & \qquad\frac{27}{8}\rho^{2}+\rho\qquad & -\frac{93}{128}\rho^{2} & -\frac{130}{256}\rho^{2}\\
-\frac{1}{16}\rho^{2} & -\frac{7}{32}\rho^{2} & -\frac{29}{64}\rho^{2} & -\frac{93}{128}\rho^{2} & \qquad\frac{251}{64}\rho^{2}+\rho & \frac{386}{512}\rho^{2}\\
-\frac{1}{32}\rho^{2} & -\frac{8}{64}\rho^{2} & -\frac{37}{128}\rho^{2} & -\frac{130}{256}\rho^{2} & \frac{386}{512}\rho^{2} & \qquad\frac{565}{128}\rho^{2}+\rho
\end{array}\right),\label{cov example}
\eea
where we further recall that due to the embedding property the covariance matrices of homogenouse ASIPs of size $n=1,...,5$ are all contained in $\mathbf{Cov}\left[\mathbf{X}(6)\right]$ as sub-matrices. 

Elements on the main diagonal of the covariance matrix in \eref{cov example} are the variances in the site occupancies. These have previously been computed in \cite{ASIP-4} and are given by \eref{variance}. The non-diagonal elements of the covariance matrix are a measure for the strength of the correlations between pairs of sites in the system. A few simple observations can be made with respect to the latter. First, note that the covariance of the first site with site $j>1$ is given by  
\begin{equation}
    \mathbf{{\color{black}\left(\mathbf{Cov}\right)_{\mathit{\mathrm{1},j}}=\mathrm{-\frac{2}{2^{j}}}}}\rho^{2}. \label{First row cor}
\end{equation}
Next, note that the covariance of the second site with site $j>2$ is given by
\begin{equation}
    \mathbf{\left(\mathbf{Cov}\right)_{\mathit{\mathrm{2},j}}=\mathrm{-\frac{j+2}{2^{j}}}}\rho^{2}.\label{Second row cor}
\end{equation}
The covariance of the third site with site $j>3$ is given by
\begin{equation}
 \mathbf{\left(\mathbf{Cov}\right)_{\mathit{\mathrm{3},j}}=-\mathrm{\frac{\frac{1}{4}j^{2}+\frac{5}{4}j+2}{2^{j}}\rho^{2}}.}   
\end{equation}
Continuing in this manner, we find that the covariance of site $i$ with site $j>i$ is given by 
\bea 
\mathbf{\left(\mathbf{Cov}\right)_{\mathit{\mathrm{i},j}}}=\frac{P_i(j)}{2^j}\rho^{2},
\label{cov_polynomial}
\eea 
where 
\bea
&\mathit{P_{\mathrm{1}}(j)}=-2 \nonumber\\
&\mathit{\mathit{P_{\mathrm{2}}(j)}}=-j-2 \nonumber \\
&\mathit{\mathit{P_{\mathrm{3}}(j)}}=-\frac{1}{4}j^{2}-\frac{5}{4}j-2\nonumber \\
&\mathit{\mathit{P_{\mathrm{4}}(j)}}=-\frac{1}{24}j^{3}-\frac{3}{8}j^{2}-\frac{4}{3}j-2 \nonumber\\
&\mathit{\mathit{P_{\mathrm{5}}(j)}}=-\frac{1}{192}j^{4}-\frac{7}{96}j^{3}-\frac{83}{192}j^{2}-\frac{131}{96}j-2 \nonumber\\
&\mathit{\mathit{P_{6}(j)}}=-\frac{1}{1920}j^{5}-\frac{1}{96}j^{4}-\frac{35}{384}j^{3}-\frac{11}{24}j^{2}-\frac{661}{480}j-2\nonumber\\
&\vdots
\label{cov polynomials examples}
\eea

The polynomials given in \eref{cov polynomials examples} are the first in an infinite sequence of polynomials. Observe the difference between every pair of consecutive polynomials 
\bea
&\Delta_{1}(j)=\mathit{\mathit{P_{\mathrm{2}}(j)}}-\mathit{P_{\mathrm{1}}(j)}=-j\nonumber\\
&\Delta_{2}(j)=\mathit{\mathit{P_{\mathrm{3}}(j)}}-\mathit{P_{\mathrm{2}}(j)}=-\frac{1}{4}j(j+1)\nonumber\\
&\Delta_{3}(j)=\mathit{\mathit{P_{\mathrm{4}}(j)}}-\mathit{P_{\mathrm{3}}(j)}=-\frac{1}{24}j(j+1)(j+2)\nonumber\\
&\Delta_{4}(j)=\mathit{\mathit{P_{\mathrm{\mathrm{5}}}(j)}}-\mathit{P_{\mathrm{4}}(j)}=-\frac{1}{192}j(j+1)(j+2)(j+3)\nonumber\\
&\Delta_{5}(j)=\mathit{\mathit{P_{\mathrm{6}}(j)}}-\mathit{P_{\mathrm{5}}(j)}=-\frac{1}{1920}j(j+1)(j+2)(j+3)(j+4).\nonumber\\
&\vdots
\label{cov differences}
\eea
We thus find that the polynomial sequence in \eref{cov polynomials examples} is a difference sequence whose differences are of the following form ($k>1$)
\begin{equation}
    \Delta_{k}(j)=-\mathrm{\frac{\mathrm{\mathit{j^{\mathrm{(k)}}}}}{2^{k-1}k!}}, \label{differences}
\end{equation}
where $\mathrm{\mathit{j^{\mathrm{(k)}}}}=j(j+1)\cdots(j+k-1)$ is the pochhammer polynomial \cite{Mathematical functions} and  $\mathrm{2^{k-1}k!}$ is the OEIS sequence A002866. 
Setting $i<j$, and summing the differences, we obtain a formula for the $P_{i}(j)$ polynomials  
\begin{equation}
P_{i}(j)=-2-\mathrm{\sum_{k=1}^{i-1}}\mathrm{\frac{\mathrm{\mathit{j^{\mathrm{(k)}}}}}{2^{k-1}k!}}=2^{1-i} \binom{i+j-1}{i} \, _2F_1\left(1,i+j;1+i;\frac{1}{2}\right)-2^{j+1}.\label{cov polynomials}
\end{equation}
Combining Eqs. (\ref{variance}) and (\ref{cov polynomials}) and utilizing the symmetry of the covariance matrix we obtain 
\begin{equation}
\left(\mathbf{Cov}\left[\mathbf{X}(n)\right]\right)_{i,j}=\begin{cases}
\left(\frac{1}{2^{i+j-1}}\left(\begin{array}{c}
j+i-1\\
i
\end{array}\right){}_{2}F_{1}\left(1,i+j,1+i;\frac{1}{2}\right)-2\right)\rho^{2} & j>i\\
\\\left(\frac{1}{2^{i+j-1}}\left(\begin{array}{c}
j+i-1\\
j
\end{array}\right){}_{2}F_{1}\left(1,i+j,1+j;\frac{1}{2}\right)-2\right)\rho^{2} & i>j\\\\
\rho+\left(\frac{4\Gamma(i+1/2)}{\sqrt{\pi}\Gamma(i)}-1\right)\rho^{2} & i=j
\end{cases}\label{homo' correlations}
\end{equation}
The result in \eref{homogeneous correlations} follows immediately.

\newpage

\section{CONCLUSIONS}
\noindent In this paper, we studied occupancy correlations in the ASIP. We obtained evolution equations for these correlations and introduced an iterative scheme for their computation in the steady-state. Using this scheme, we solved for the steady-state correlations of general ASIPs and obtained a closed-form expression for the correlations of homogeneous ASIPs. In small ASIPs, Monte-Carlo simulations provide reliable estimates for occupancy correlations which allowed us to validate our analytical results. As the latter also extend to ASIP systems of arbitrary size --- the results we have obtained provide a comprehensive description of spatial correlations in the ASIP which would be virtually impossible to get using brute-force Monte-Carlo simulations.    

\begin{figure}[t]
\includegraphics[width=16cm]{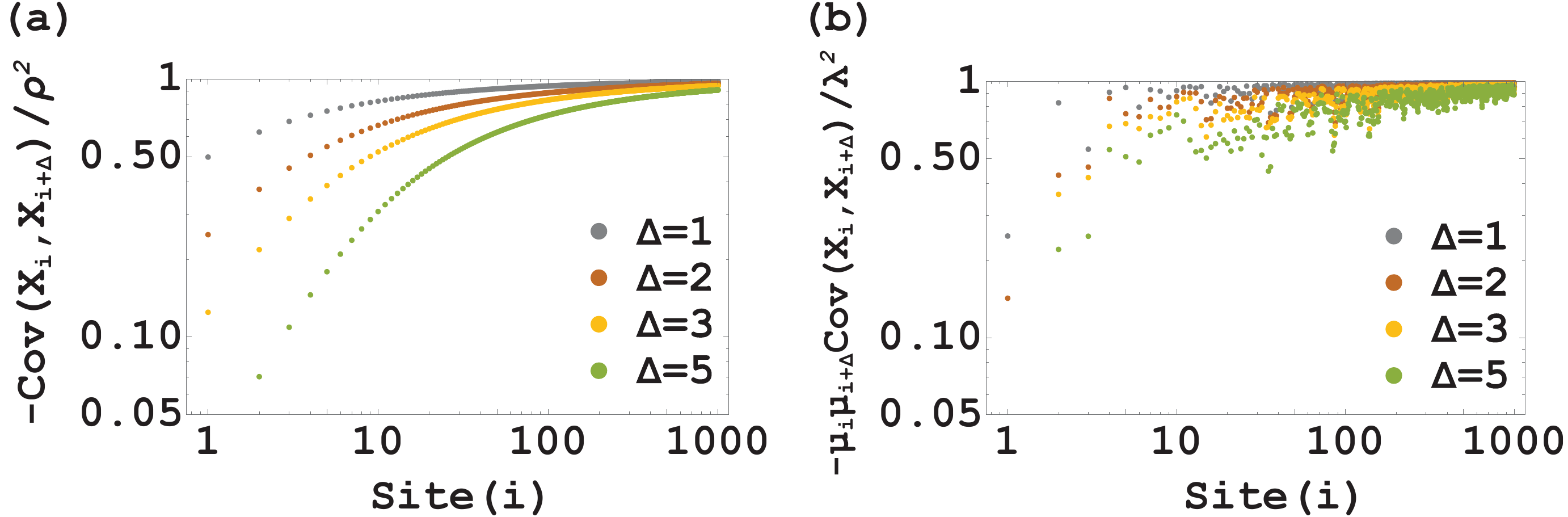}\caption{In the ASIP, the covariance between the occupancies of sites $i$ and site $i+\Delta$ tends to minus the product of their mean occupancies as $i\to\infty$. We thus have $-\mathbf{Cov}(X_i,X_{i+\Delta})\simeq\lambda^2/\mu_i\mu_{i+\Delta}=\langle X_i \rangle \langle X_{i+\Delta} \rangle \neq \langle X_iX_{i+\Delta} \rangle \simeq 0$, which suggests strong occupancy correlations. Plots are made using  Eqs. (\ref{general correlations}-\ref{homogeneous correlations}).}
\label{Fig4}
\end{figure} 

In Figs. (\ref{Fig2}) and (\ref{Fig3}), we utilized the results in Eqs. (\ref{general correlations}-\ref{homogeneous correlations}) to explore how occupancy correlations between sites $i$ and $j$ decay with the distance $\Delta=j-i$ between them $(i<j)$. For large $\Delta$, this decay was seen to be roughly exponential; but Eqs.  (\ref{cov_polynomial}-\ref{cov polynomials examples}) assert that the decay is further modulated by a polynomial of degree $i-1$, i.e., one less than the position of the upstream site $i$ from which the distance between the two sites is measured. 

Fixing a distance $\Delta$, one could also ask about the correlation between sites $i$ and $j=i+\Delta$ as this pair is pushed downstream together, i.e., as $i,j\to \infty$. Utilizing Eqs. (\ref{general correlations}-\ref{homogeneous correlations}), we find that for $i,j\gg1$,  $\mathbf{\left(\mathbf{Cov}\right)_{\mathit{\mathrm{i},j}}}\simeq-\lambda^2/\mu_i\mu_j$,  which for homogeneous ASIPs boils down to $\mathbf{\left(\mathbf{Cov}\right)_{\mathit{\mathrm{i},j}}}\simeq-\rho^2$ (Fig. 4). This result asserts that in this limit $\langle X_iX_j \rangle \simeq 0$, which in turn means that two ASIP sites that are located far away downstream are strongly correlated as they are rarely occupied simultaneously. 

To better understand the steady-state structure of the ASIP one should go beyond the correlation functions considered herein. While the full steady-state distribution of the ASIP still poses a formidable challenge, calculating the the joint steady-state distribution of site pairs could serve as an interim goal.  Specifically, one would like to know $\text{Pr}(X_i=k_i,X_j=k_j)$, i.e., the probability to simultaneously find $k_i$ particles in site $i$ and $k_j$ particles in site $j$ ($k_i,k_j=0,1,2,3,\cdots$). Generalizing the results obtained herein to ASIP systems with arbitrary, i.e., not necessarily non-exponential, gate opening times is also of interest \cite{ASIP-5}. Doing so will require methods that go beyond those that were used above.    

\section{Acknowledgments}
Shlomi Reuveni acknowledges support from the Azrieli Foundation and from the Raymond and Beverly Sackler Center for Computational Molecular and Materials Science at Tel Aviv University. S. Reuveni and O. Lauber, thank Yuval Scher for reading and commenting on an early version of this manuscript.

\clearpage

\section{APPENDIX}

\subsection{DERIVATION OF EQ. (\ref{variance})}

\noindent In this appendix, we derive Eq. (\ref{variance}) based on a known result for the factorial moments of ${\color{black}\mathrm{X}_{\mathrm{k}}}$. In an homogeneous ASIP, the factorial moments are given by (Eq. ({73}) in \cite{ASIP-4}) \begin{equation}
\left\langle \mathrm{X_{k}(X_{k}-1)\cdots(X_{k}-\mathit{l}+1)}\right\rangle =\frac{2^{l}\rho^{l}\Gamma(1+l/2)\Gamma(k+l/2-1/2)}{\sqrt{\pi}\Gamma(k)},\label{factorial moments}
\end{equation}
where ${\color{black}k=1,\ldots,n}$ and ${\color{black}l=1,2,3,\ldots}$. In order to calculate the variance of ${\color{black}\mathrm{X}_{\mathrm{k}}}$, we first calculate the first two factorial moments of $\mathrm{X}_{\mathrm{k}}$. Setting $l=1$ and $l=2$ in Eq. (\ref{factorial moments}), we find
\begin{equation}
\left\langle \mathrm{X_{k}}\right\rangle =\frac{2\rho\Gamma(3/2)\Gamma(k)}{\sqrt{\pi}\Gamma(k)}=\frac{2\rho\frac{\sqrt{\pi}}{2}}{\sqrt{\pi}}=\rho,\label{first factorial moment}
\end{equation}
and
\begin{equation}
\left\langle \mathrm{X_{k}(X_{k}-1)}\right\rangle =\frac{4\rho^{2}\Gamma(2)\Gamma(k+1/2)}{\sqrt{\pi}\Gamma(k)}=\frac{4\rho^{2}\Gamma(k+1/2)}{\sqrt{\pi}\Gamma(k)}.\label{second factorial moment}
\end{equation}
Combining Eqs. (\ref{first factorial moment}) and (\ref{second factorial moment}), we see that the second moment is given by
\begin{equation}
\left\langle \mathrm{X}_{k}^{2}\right\rangle =\left\langle \mathrm{X_{k}}\right\rangle +\left\langle \mathrm{X_{k}(X_{k}-1)}\right\rangle =\rho+\frac{4\rho^{2}\Gamma(k+1/2)}{\sqrt{\pi}\Gamma(k)}.\label{second moment of Xk}
\end{equation}
The variance can now be obtained by combining Eqs. (\ref{first factorial moment}) and (\ref{second moment of Xk}) 
\begin{equation}
{\color{black}\mathbf{Var[\mathbf{\mathrm{\mathrm{X}}_{\mathrm{k}}}]=\left\langle \mathrm{X}_{\mathrm{k}}^{\mathrm{2}}\right\rangle -\left\langle \mathrm{X_{k}}\right\rangle ^{\mathrm{2}}}}{\color{black}=\rho+\left(\frac{4\Gamma(k+1/2)}{\sqrt{\pi}\Gamma(k)}-1\right)\rho^{2}},
\end{equation}
which is identical to Eq. (\ref{variance}).


\newpage
\begin{thebibliography}{1}

\bibitem{ASIP-1}S. Reuveni, I. Eliazar and U. Yechiali, Asymmetric Inclusion Process. \emph{Phys. Rev. E.} \textbf{84}, 041101, (2011). 

\bibitem{ASIP-2}S. Reuveni, I. Eliazar and U. Yechiali, The Asymmetric Inclusion Process: A Showcase of Complexity. \emph{Phys. Rev. Lett.} \textbf{109}, 020603, (2012).

\bibitem{R. R. P. Jackson 1}R. R. P. Jackson, Queueing systems with phase-type service. \emph{Operational Research Quarterly,} \textbf{5} (4), 109-120, (1954).

\bibitem{R. R. P. Jackson 2}R. R. P. Jackson, Random Queueing Processes with Phase-Type Service. \emph{Journal of the Royal Statistical Society} \emph{Series B (Methodological)}, \textbf{18}, 1, 129-132, (1956).

\bibitem{Harchol-Balter}Harchol-Balter, M., 2013. Performance modeling and design of computer systems: queueing theory in action. Cambridge University Press.

\bibitem{Kendall}D. G. Kendall, Stochastic Processes Occurring in the Theory of Queues and their Analysis by the Method of the Imbedded Markov Chain. \emph{The Annals of Mathematical Statistics,} \textbf{24},
3, 338, (1953).

\bibitem{Derrida1}B. Derrida, E. Domany and D. Mukamel. An exact solution of the one dimensional asymmetric exclusion model with open boundaries. \textit{Journal of Statistical Physics,} \textbf{69}, 667, (1992).

\bibitem{Golinelli}O. Golinelli and K. Mallick, The asymmetric simple exclusion process: an integrable model for non-equilibrium statistical mechanics. \emph{J. Phys. A}: Math. Gen. \textbf{39}, 12679, (2006). 

\bibitem{Derrida2}B. Derrida, Non-equilibrium steady-states: fluctuations and large deviations of the density and of the current.\textit{ J. Stat. Mech.,} P07023, (2007).

\bibitem{Blythe}R. A. Blythe and M. R., Evans, Nonequilibrium steady-states of matrix-product form: a solver’s guide. \textit{J. Phys. A: Math. Theor.,} \textbf{40}, R333-R441, (2007). 

\bibitem{MacDonald}C. T. MacDonald, J. H. Gibbs and A. C. Pipkin, Kinetics of biopolymerization on nucleic acid templates. \textit{Biopolymers,} \textbf{6}, 1, (1968).

\bibitem{Spitzer}F. Spitzer, Interaction of Markov processes. \emph{Adv.Math.,} \textbf{5}, 246, (1970).

\bibitem{Heckmann}K. Heckmann, Single file diffusion Passive Permeability of Cell Membranes. \textit{Biomembranes,} \textbf{3}, 127, (1972).

\bibitem{Levitt}D. G. Levitt, Dynamics of a single-file pore: Non-Fickian behavior. \emph{Phys. Rev. A,} \textbf{8}, 3050, (1973).

\bibitem{Richards}P. M. Richards, Theory of one-dimensional hopping conductivity and diffusion. \emph{Phys. Rev. B,} \textbf{16}, 1393, (1977).

\bibitem{Widom}B. Widom, J. L. Viovy and A. D. Defontaines, Repton model of gel electrophoresis and diffusion. \emph{J. Physique I,} \textbf{1}, 1759, (1991).

\bibitem{Schreckenberg}M. Schreckenberg and D. E. Wolf (ed), Traffic and Granular Flow, New York: Springer, (1998).

\bibitem{Reuveni}S. Reuveni, I. Meilijson, M. Kupiec, E. Ruppin and T. Tuller, Genome-Scale Analysis of Translation Elongation with a Ribosome Flow Mode. \emph{PLoS Computational Biology,} \textbf{7}(9), e1002127, (2011).

\bibitem{Shaw} L. B. Shaw, R.K. Zia and K.H. Lee, Totally asymmetric exclusion process with extended objects: a model for protein synthesis. \emph{Phys Rev E,} \textbf{68}, 021910, (2003).

\bibitem{Halpin}T. Halpin-Healy and Y. C. Zhang, Kinetic roughening phenomena, stochastic growth, directed polymers and all that. \emph{Phys. Rep.,} \textbf{254}, 215, (1995).

\bibitem{Krug}J. Krug, Origins of scale invariance in growth processes. \emph{Adv. Phys.,} \textbf{46}, 139, (1997).

\bibitem{Bundschuh}R. Bundschuh, Asymmetric exclusion process and extremal statistics of random sequences, \emph{Phys. Rev. E,} \textbf{65}, 031911, (2002).

\bibitem{Klumpp}S. Klumpp and R. Lipowsky, Traffic of molecular motors through tube-like compartments, \emph{J. Stat. Phys.,} \textbf{113}, 233, (2003).

\bibitem{Oshanin1}S. F. Burlatsky, G. S. Oshanin, A. V. Mogutov and M. Moreau, Directed walk in a one-dimensional lattice gas. \emph{Physics Letters A, }\textbf{166}, 230-234, (1992).

\bibitem{Oshanin2}S. F. Burlatsky, G. Oshanin, M. Moreau and W. P. Reinhardt, Motion of a driven tracer particle in a one-dimensional symmetric lattice gas. \emph{Phys Rev E,} \textbf{54}, 3165-3172, (1996).

\bibitem{Oshanin3}O. Bénichou, A. M. Cazabat, J. De Coninck, M. Moreau and G. Oshanin, Stokes Formula and Density Perturbances for Driven Tracer Diffusion in an Adsorbed Monolayer. \emph{Phys. Rev. Lett.,}
\textbf{84}, 511-514, (2000).

\bibitem{Monasterio}C. M. Monasterio and Gleb Oshanin, Bias and bath mediated pairing of particles driven through a quiescent medium. \emph{Soft Matter,} \textbf{7}, 993-1000, (2011).

\bibitem{Derrida3}B. Derrida, M.R. Evans, V. Hakim and V. Pasquier, Exact solution of a 1d asymmetric exclusion model using a matrix formulation. \emph{J. Phys. A,} \textbf{26}, 1493-1517, (1993).

\bibitem{ASIP-3}S. Reuveni, I. Eliazar and U. Yechiali, Limit Laws for the Asymmetric Inclusion Process. \emph{Phys. Rev. E.} \textbf{86},
061133, (2012).

\bibitem{ASIP-4}Reuveni, S., Hirschberg, O., Eliazar, I. and Yechiali, U., 2014. Occupation probabilities and fluctuations in the asymmetric simple inclusion process. Physical Review E, 89(4), p.042109.

\bibitem{ASIP-5}Boxma, O., Kella, O. and Yechiali, U., 2016. An ASIP model with general gate opening intervals. Queueing Systems, 84(1-2), pp.1-20.

\bibitem{Not-ASIP1}Grosskinsky, S., Redig, F. and Vafayi, K., 2011. Condensation in the inclusion process and related models. Journal of Statistical Physics, 142(5), pp.952-974.

\bibitem{Not-ASIP2}Cao, J., Chleboun, P. and Grosskinsky, S., 2014. Dynamics of condensation in the totally asymmetric inclusion process. Journal of Statistical Physics, 155(3), pp.523-543.

\bibitem{Neuts}M. F. Neuts, The Busy Period of a Queue with Batch Service. \emph{Operations Research,} \textbf{13}, 815-819, (1965).

\bibitem{Kaspi}H. Kaspi, O. Kella and D. Perry, Dam processes with state dependent batch sizes and intermittent production processes with state dependent rates. \emph{Queueing Systems: Theory and Applications,}
\textbf{24}, 37-57, (1997).

\bibitem{Boxma}O. Boxma, D. Perry, W. Stadje and S. Zacks, A Markovian growth-collapse model. \emph{Advances in Applied Probabbility,} \textbf{38}, 221-243, (2006). 

\bibitem{Kella}O. Kella, On growth collapse processes with stationary structure and their shot-noise counterparts. \emph{Journal of Applied Probability,} \textbf{46}, 363-371, (2009). 

\bibitem{Martin}B. J. Martin, Batch queues, reversibility and first-passage percolation. \emph{Queueing Systems: Theory and Applications}, \textbf{62},
411-427, (2009). 

\bibitem{sandpile}P. Bak, How nature works: the science of self organized criticality. \emph{Copernicus}, (1996).

\bibitem{Rozman}M. G. Rozman, M. Urbach, J. Klafter and F. J. Elmer, Atomic scale friction and different phases of motion of embedded molecular systems. \emph{J. Phys.} \emph{Chem. B,} \textbf{102}, 7924-7930, (1998).

\bibitem{Carlson}J. M. Carlson, J. S. Langer and B. E. Shaw, Dynamics of earthquake faults. \emph{Rev. Mod. Phys.,} \textbf{66}, 657-670, (1994).

\bibitem{Eliazar1}I. Eliazar and J. Klafter, A growth-collapse model: Lévy inflow, geometric crashes, and generalized Ornstein-Uhlenbeck dynamics. \emph{Physica A,} \textbf{334}, 1-21, (2004).

\bibitem{Eliazar2}I. Eliazar and J. Klafter, Stochastic Ornstein-Uhlenbeck capacitors. \emph{Journal of Statistical Physics,} \textbf{118}, 177-198, (2005).

\bibitem{Eliazar3}I. Eliazar and J. Klafter, Growth-collapse and decay-surge evolutions, and geometric Langevin equations. \emph{Physica A,} \textbf{367}, 106-128, (2006).

\bibitem{Smoluchowski}M. Smoluchowski, Versuch einer mathematischen Theorie der Koagulationskinetik kolloider Lösungen. \emph{Z. phys. Chem.,} \textbf{92}, (1917).

\bibitem{Sokolov}I. M. Sokolov, S. B. Yuste, J. J. Ruiz-Lorenzo and K. Lindenberg. Mean field model of coagulation and annihilation reactions in a medium of quenched traps: Subdiffusion. \emph{Phys. Rev. E,}
\textbf{80}, 051114, (2009).

\bibitem{Lindenberg}S. B. Yuste, J. J. Ruiz-Lorenzo and K. Lindenberg. Coagulation reactions in low dimensions: Revisiting subdiffusive A+A reactions in one dimension. \emph{Phys. Rev. E,} \textbf{80}, 051114, (2009).

\bibitem{Kinetic View}P. L. Krapivsky, S. Redner and E. Ben-Naim. A kinetic view of statistical physics. Cambridge University Press, Cambridge, UK, 2010.

\bibitem{Coalescence Process}D. Ben-Avraham. The coalescence process, $A+A\rightarrow A$, and the method of interparticle distribution functions. In V. Privman, editor, Nonequilibrium Statistical Mechanics in One Dimension, pages 29--50. Cambridge University Press, Cambridge, UK, 2005.

\bibitem{zero-range}Evans, M.R. and Hanney, T., 2005. Nonequilibrium statistical mechanics of the zero-range process and related models. Journal of Physics A: Mathematical and General, 38(19), p.R195.

\bibitem{Catalan's Trapezoids}Reuveni, S., 2014. CATALAN'S TRAPEZOIDS. Probability in the Engineering and Informational Sciences, 28(3), pp.353-361. 

\bibitem{Mathematical functions}Abramowitz, M. and Stegun, I.A., 1965. Handbook of mathematical functions: with formulas, graphs, and mathematical tables (Vol. 55). Courier Corporation.
\end{thebibliography}
\end{document}